\documentclass[%
reprint,
 amsmath,amssymb,
 aps,
]{revtex4-2}

\usepackage{graphicx}
\usepackage{dcolumn}

\usepackage{appendix}
\usepackage{color}
\usepackage{xcolor}
\usepackage{bm}

\begin{document}


\title{ \color{black!40!black} Role of interactions in non-equilibrium transformations }

\author{Maria Rose}
 \affiliation{School of Pure and Applied Physics, Mahatma Gandhi University, Kottayam, India }
\author{Sreekanth K Manikandan}%
 \email{sreekm@stanford.edu}
 \thanks{In this study, part of the work was conducted by the author while employed at NORDITA, and the remaining part was carried out while the author was affiliated with Stanford University.}
\affiliation{NORDITA, KTH Royal Institute of Technology and
Stockholm University, Roslagstullsbacken 23, 10691 Stockholm, Sweden}
\affiliation{Department of Chemistry, Stanford University, Stanford, CA, USA 94305
}%

\date{\today}

\begin{abstract}
For arbitrary non-equilibrium transformations in complex systems, we show that the distance between the current state and a target state can be decomposed into two terms: one corresponding to an \textit{independent} estimate of the distance, and another corresponding to interactions, quantified using the relative mutual information between the variables. This decomposition is a special case of a more general decomposition involving successive orders of correlation or interactions among the degrees of freedom of the system. {\color{black!0!black} To illustrate its practical significance, we study the thermal relaxation of
two interacting, optically trapped colloidal particles, where increasing pairwise interaction strength is shown to prolong the longevity of the time-dependent non-equilibrium state. Additionally, we study a system with both pairwise and triplet interactions, where our approach identifies their distinct contributions to the transformation.} In more general setups where it is possible to control the strength of different orders of interactions, our findings provide a way to disentangle their effects and identify interactions that facilitate the transformation.
\end{abstract}

\maketitle

\section{Introduction}
A broad range of microscopic non-equilibrium processes are time-dependent, where the state of the system, described in terms of probability distributions, changes as a function of time. Examples include the thermal relaxation of systems prepared in an arbitrary initial state \cite{dattagupta2012relaxation}, self-assembly of biological molecules \cite{whitesides2002self,pollard_cellular_2003,mauro2014self}, protein folding \cite{dobson2003protein,creighton1990protein}, several single-molecule experiments \cite{ritort2006single,ciliberto2017experiments}, and microscopic devices that are time-dependently controlled \cite{pop2010energy,bergfield2013forty,martinez2016brownian}. In all these cases, the trajectory of the system progresses through a series of states, influenced by interactions among the different degrees of freedom of the system, with the environment, and external controls/feedbacks \cite{sagawa2012fluctuation,barato2014unifying}.

Several recent studies have tried to identify governing principles for such processes in terms of the distance between the initial and final states of the system, the time taken for the transformation, and the associated thermodynamic costs. These include the refinements of the Second Law \cite{aurell2012refined,Lutz,kim2021information,nakazato2021geometrical}, optimal connections \cite{ito2023geometric,chennakesavalu2023unified,rotskoff2015optimal}, speed limits \cite{shiraishi2018speed,van2023topological,yoshimura2021thermodynamic,funo2019speed} as well as their trade-offs with the entropic costs \cite{lee2022speed,falasco2020dissipation,van2023thermodynamic,yoshimura2021thermodynamic,kuznets2021dissipation}. However, the fundamental effects of interactions among the different degrees of freedom of the system, on the distance or time taken for non-equilibrium transformations are relatively less understood.

In a recent development, Refs. \cite{arrowoftime,lynn2022emergence} made significant progress in this direction. They demonstrated that in systems with multiple degrees of freedom and having multi-partite dynamics, the estimate of \textit{irreversibility} in a non-equilibrium steady state can be decomposed into contributions from individual variables, and a series of non-negative contributions from correlations among
variable pairs, triplets, and higher-order combinations. Their proof is based on representing irreversibility as a Kullback-Leibler divergence, which measures the relative likelihood of trajectories over their time-reversed counterparts. 

In general, the Kullback-Leibler divergence quantifies the distance between any two probability distributions, and it has recently gained renewed interest in studying non-equilibrium transformations and control of microscopic systems \cite{ito2018stochastic,aurell2011optimal,nakazato2021geometrical,chennakesavalu2023adaptive}. In certain cases, it also provides estimates of the thermodynamic cost of the process \cite{shiraishi2019information,chetrite2021metastable,sagawa2012fluctuation}. Hence, understanding how this distance function depends on interactions is crucial, as it enables the optimization of processes based on interactions, and the design of more efficient and reliable non-equilibrium controls.

Here we address this problem by implementing a decomposition of the Kullback-Leibler divergence. This decomposition primarily consists of two terms: one corresponding to an \textit{independent} estimate of the distance, representing hypothetical marginal processes which are non-interacting, and another corresponding to interactions, quantified using the relative mutual information between the variables. This decomposition is further shown to arise from a previously known decomposition of the joint distribution involving successive orders of correlation or interactions among the system's degrees of freedom \cite{mcclendon2012comparing,galas2017expansion,tritchler2011information}. Crucially this decomposition is not limited to multi-partite systems. Applying the decomposition to an interacting pair of colloids that undergo thermal relaxation, we find that increasing the strength of pairwise interactions generically increases the distance between the current state and the target state, prolonging the longevity of the time-dependent non-equilibrium state. {\color{black!0!black} Additionally, in a three-variable case with pairwise and triplet interactions, our approach isolates their distinct contributions to the transformation process. For both systems, we also discuss the effects of external non-conservative forces.} In more general setups, where it is possible to control the strength of different orders of interactions, our results can potentially be used to separate out their effects on the transformation process.
\section{Results}
We begin by considering a system whose state is described using the variable ${\bm x}_t \in \mathbb{R}^N$, and probability distribution $P({\bm x}_t)$. We have dropped the explicit dependence on $t$ for simplicity of notation. Note that one of the elements of vector ${\bm x}_t$ can also be an external control or a feedback protocol. Let us now consider a scenario where the probability distribution $P({\bm x}_t)$ dynamically evolves from an initial distribution $P_i({\bm x}_{t_i})$ to a final / target distribution $P_f({\bm x}_{t_f})$ in a time-dependent manner. At any given time $t$, the distance of the instantaneous distribution $P({\bm x}_t)$ to the target distribution can be computed in terms of the Kullback-Leibler (KL) divergence between the two distributions as \cite{lu2017nonequilibrium}, 
\begin{align}
\begin{split}
   D_{\rm KL}(P({\bm x}_t) \vert \vert P_f({\bm x}_{t})) &= \int_{{\bm x}_t} P( {\bm x}_t) \log \frac{P( {\bm x}_t)}{P_f({\bm x}_{t})}.
\end{split}
\end{align}
Next, assume we know the marginal distributions, $P_m^i(x_t^i) = \int_{{\bm x}_{-i}}P( {\bm x}_t)$,
where ${\bm x}_{-i}$ corresponds to all variables except $x_t^i$.
One can obtain an \textit{independent} distance in terms of these marginals as, 
\begin{align}
\begin{split}
\label{eq:Dind}
        D^i_{\rm{Ind}}  = \int_{x_t^i}P_m^i(x_t^i)\log \frac{P_m^i(x_t^i)}{P_{f,m}^i(x_t^i)}.
\end{split}
\end{align}
The sum of the independent distances over all variables, $D_{\rm Ind.} = \sum_i D^i_{\rm{Ind}}$, provides an estimate of the distance that one would have got if the variables were independently measured. By examining the difference $D - D_{\rm Ind.}$, we find,
\begin{align}
\label{eq:KLinteraction}
\begin{split}
    D &- \sum_i D^i_{\rm{Ind}} \\
    &=\int_{{\bm x}_t} P( {\bm x}_t) \left[\log \frac{P( {\bm x}_t)}{\prod_i P_m^i(x_t^i)} - \log \frac{P_f( {\bm x}_t)}{\prod_i P_{m,f}^i(x_t^i)}\right]\\
    &= I({\bm x}_t) - I_f^\prime ({\bm x}_t),
    \end{split}
\end{align}
where $I({\bm x}_t)$ is the mutual information of the current state, generalized to $N$ variables (also referred to as the total correlation \cite{watanabe1960information}), and $I_f^\prime({\bm x}_t)$ is the cross mutual information of the target state, where the average is computed with respect to the current state. 

Eq.\ \eqref{eq:KLinteraction} is our first key observation: the distance between any two distributions can be decomposed into two terms: a term coming from the marginal probabilities and another coming from interactions between the local variables, {\it i.e.}, 
\begin{align}
\label{eq:decomposition0}
    D =  D_{\rm Ind.} + D_{\rm Int.},
\end{align}
where $D_{\rm Int.} \equiv I({\bm x}_t) - I_f^\prime ({\bm x}_t)$, appears as the relative mutual information between the current state and the target state. Note that the sign of this interaction term could be positive or negative, depending on the choice of the final distribution and the nature of interactions. Eq.\ \eqref{eq:KLinteraction} also has a simple information theoretic interpretation: Interactions contribute to the distance only if the mutual information of the current state differs from the cross mutual information of the target state. This means, there could be instances where accurate distance measurements can be solely obtained from the marginal statistics, even when the local variables are correlated.

In a similar spirit, one can argue that the total distance further breaks down into contributions from interactions among subsets of $k<N$ variables. However, the choice of this decomposition is not necessarily unique.  
Here we consider one such decomposition, which is due to the generalized Kirkwood superposition approximation \cite{mcclendon2012comparing,tritchler2011information,somani2009sampling, killian2007extraction,galas2017expansion}. In the following, we briefly describe it for conciseness.

Assume that we know all the $(N-1)^{\rm th}$ order marginal distributions,
\begin{align}
    P_{N-1}(x_1,\dots x_{N-1}) = \int _{{\bm x_t}^{-[N-1]}}P({\bm x}_t),
\end{align}
where the integration is done over the variable that is not in the subset $\lbrace x_1,\dots x_{N-1} \rbrace$. The Kirkwood superposition approximation provides an estimate to the joint probability distribution $\hat{P}_{N-1}({\bm x}_t) \simeq P({\bm x}_t)$ in terms of these marginals, as \cite{mcclendon2012comparing,tritchler2011information},
\begin{align}
\label{eq:final}
    \log \hat{P}_{N-1}({\bm x}_t) = \sum_{\alpha=1}^{N-1}(-1)^{N-\alpha+1} \log \prod_{j=1}^{C^N_\alpha} P_{\alpha}^j,
\end{align}
where the product is over all marginal densities $P_{\alpha}^j$ obtained for a subset of variables of size $\alpha \leq N-1$ { \color{black!0!black}(see Appendix A for the approximations to order 3. See also Ref.\ \cite{galas2017expansion}, where the first few terms of this approximation is derived explicitly using the Möbius inversion duality between multivariable entropies and
multivariable interaction information \cite{galas2016multivariate}, which allows a series
expansion of KL divergence in the number of interacting variables.)}. 

By successively applying the Kirkwood approximation to the RHS of Eq.\ \eqref{eq:final}, we can get an estimate of the joint distribution $P({\bm x}_t)$ in terms of marginals of any order $k < N$. We refer to the resulting $k-$th order approximation as $\hat{P}_{k}({\bm x}_t)$. In particular, for $k=1$, we will arrive at the product of single-variable marginals \cite{killian2007extraction,galas2017expansion}.{ \color{black!0!black} While lacking appropriate normalization of probability density functions for terms beyond the first order, prior studies have found meaningful applications of this approximation. These include quantifying higher-order mutual information to measure frustration \cite{Matsuda} and assessing the impact of higher-order correlations on configurational entropy changes in biologically relevant processes \cite{fenley2014correlation,killian2007extraction}. Its utility in efficiently sampling equilibrium distributions is also established \cite{mcclendon2012comparing}.} Inspired by these studies, we use the Kirkwood approximation to obtain an estimate of the distance that is accurate to $k$-th order interactions, as,
\begin{align}
\label{eq: decomp}
    D^{(k)} = \int_{{\bm x}_t} P({\bm x}_t)\log \frac{ \hat{P}_{k}({\bm x}_t)}{\hat{P}_{f,k}({\bm x}_t)}.
\end{align}
Due to the expansion in Eq.\ \eqref{eq:final}, $D^{(k)}$ is fully determined in terms of marginal probabilities upto order $k$. For $k=1$, we recover $D^{(1)} = D_{\rm Ind.}$. We can also safely define $D^{(N)} \equiv D$.  It is then natural to compare $D^{(k)}$ with $D^{(k-1)}$. If $D^{(k)} = D^{(k-1)}$, it implies that the $k$th-order dynamics is redundant, as it does not contribute to the total distance. However, if that is not the case, then the $k$th-order dynamics contribute, and we can separate the contribution as,  
\begin{align}
    D^{(k)}_{\rm Int.}=D^{(k)} -D ^{(k-1)}.
\end{align}
This yields the full decomposition of the total distance into interactions of different orders as,
\begin{align}
\label{eq:decomposition_new}
   D =  D^{(1)}_{\rm Int.}+D^{(2)}_{\rm Int.}+D^{(3)}_{\rm Int.} \cdots + D^{(N)}_{\rm Int.},
\end{align}
where $D^{(1)}_{\rm Int.} = D_{\rm Ind.}$. 

Note that the decomposition above is similar in spirit to the decomposition of irreversibility for multi-partite systems (see Refs. \cite{arrowoftime,lynn2022emergence}), breaking down the distance between two distributions into contributions from individual elements in the system, interactions between pairs of elements, interactions among triplets, and so on. However, the derivation of Eq. \eqref{eq:decomposition_new} does not assume multi-partite dynamics. Additionally, individual terms in the expansion, $D^{(k)}$, can be negative. In practice, $D^{(k)}$ can be computed from the knowledge of the full joint distribution or empirically obtained distributions, where only a collection of $k$ variables are measured simultaneously. 

\begin{figure*}
    \centering
    \includegraphics[scale = .6]{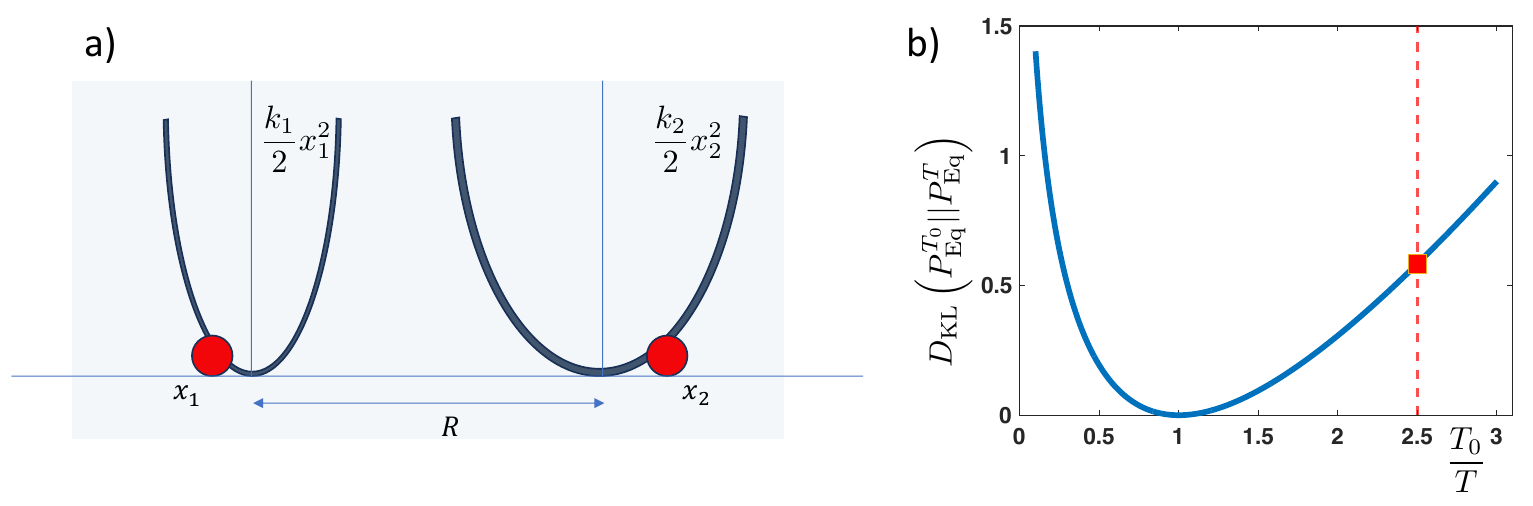}
    \caption{a) Schematics of two identical, hydrodynamically coupled colloidal particles in two spatially separated quadratic potential wells of stiffness $k_1$ and $k_2$. b) The distance between the initial equilibrium system at temperature $T_0$ and the final equilibrium system at temperature $T$. We consider a particular parameter choice $T_0/T = 2.5$, as marked. The other parameter choices are: $k_1=1;\;k_2=2;\;\gamma =1;\;\eta =1;\;k_B=1.$}
    \label{fig:schematics}
\end{figure*}

\begin{figure*}

    \centering
    \includegraphics[scale = .6]{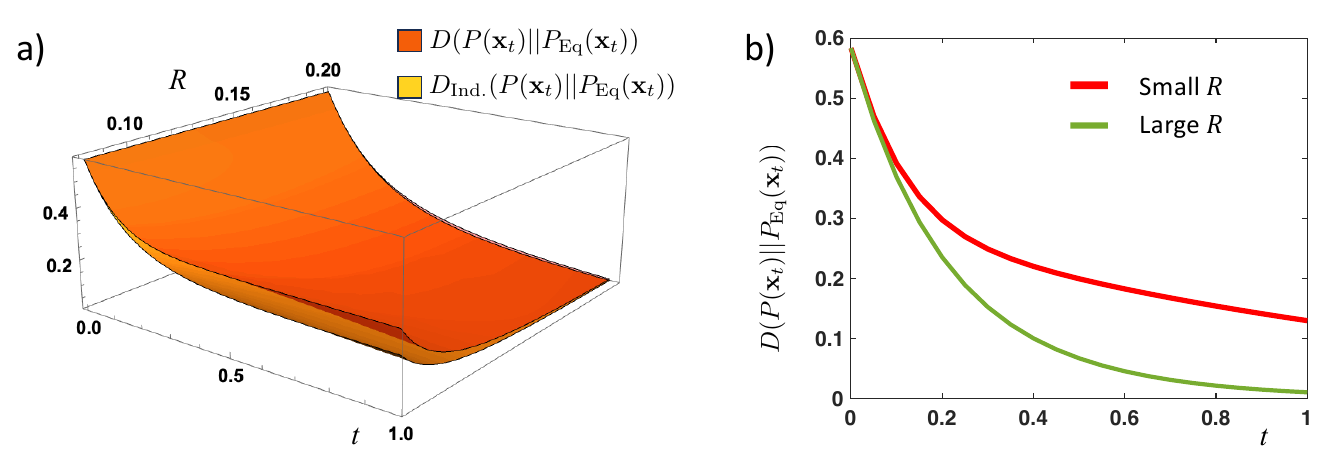}
    \caption{a) The total distance function $D \equiv D_{\rm KL}\left( P({\bm x}_t) \vert\vert P_{\rm Eq} ({\bm x}_t) \right)$ as well as the independent distance  $D_{\rm Ind.}$ for different values of $R$ and $t$, for fixed values of initial and final temperatures as well as other model parameters. We find that $D> D_{\rm Ind.}$ for all values of $R$ and $t$. b) The total distance function $D \equiv D_{\rm KL}\left( P({\bm x}_t) \vert\vert P_{\rm Eq} ({\bm x}_t) \right)$ for two values of the separation $R$ ( \textit{Top}: $R = 0.1$ and \textit{Bottom}: $R = 0.5$) between the two optical traps. We find that the system at small separation takes longer to thermalize.}
    \label{fig:Rt_dependence}
\end{figure*}

\begin{figure*}
    \centering
    \includegraphics[scale = .6]{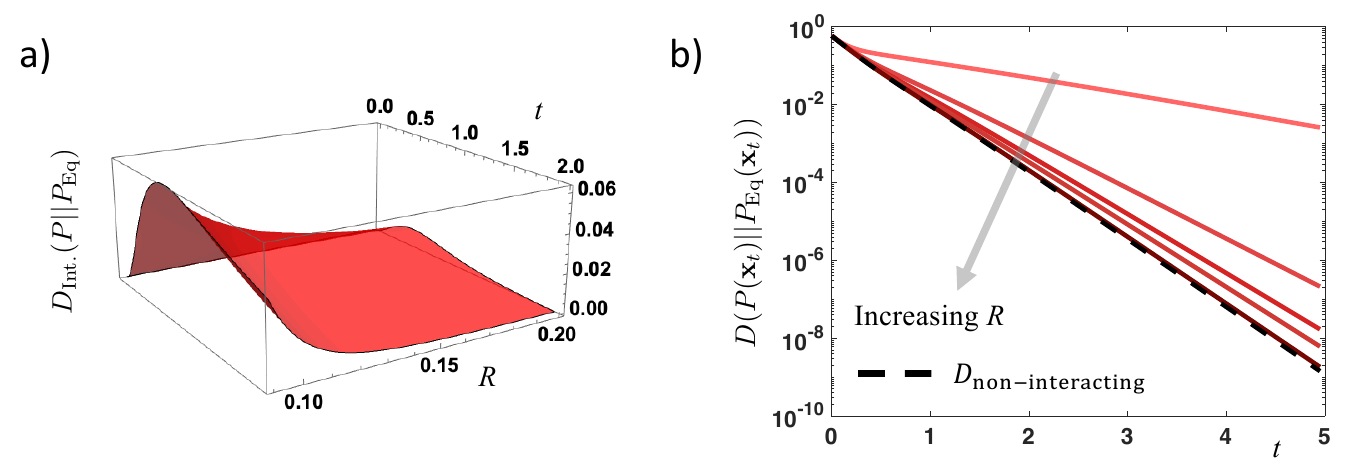}
    \caption{a) The interaction distance $D_{\rm Int.}$ for different values of time $t$ and separation $R$. We find that as we decrease $R$ and bring the two particles closer to each other, the interaction distance  $D_{\rm Int.}$ increases. b) The total distance $D$ compared with the distance computed for the non-interacting case $D_{\rm non-interacting}\equiv\lim_{R \rightarrow \infty } D $, for different values of $R$. As expected, we find that $D_{\rm non-interacting} \leq D $ for all values of $R$ and $t$, saturating the bound in the $R\rightarrow \infty$ limit. }
    \label{fig:R_dependence}
\end{figure*}

\begin{figure*}
    \centering
    \includegraphics[scale = .6]{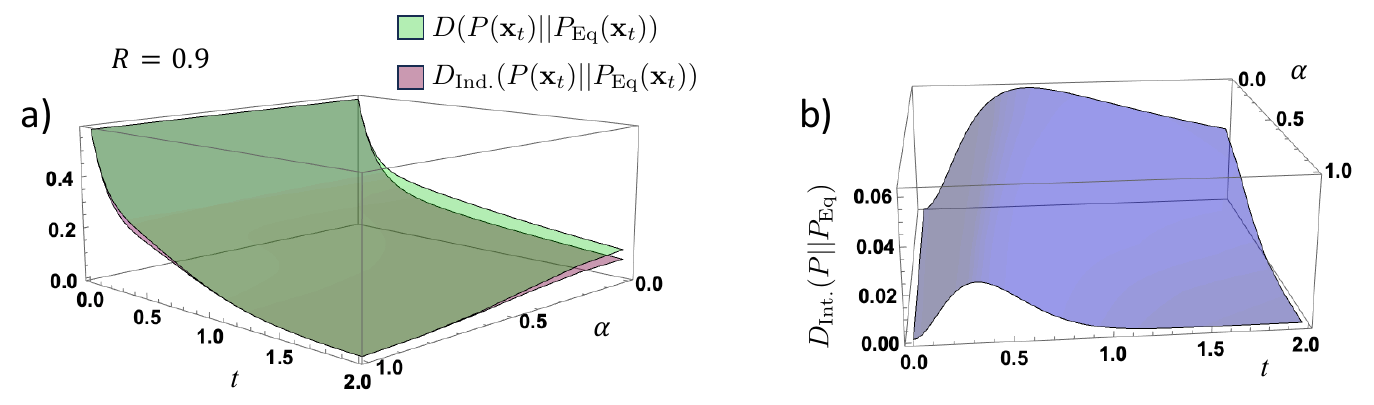}
    \caption{\color{black!0!black} a) The total distance function $D$ as well as the independent distance  $D_{\rm Ind.}$ for different values of strength of the external driving, $\alpha$ and $t$, for a fixed value of $R$ and other model parameters. b) The interaction distance $D_{\rm Int.}$ for different values of time $t$ and $\alpha$ for the parameter choice in (a). We find that all the distance functions decrease in value for any $t$ with increasing $\alpha$. The other parameter choices are: $k_1=1;\;k_2=2;\;\gamma =1;\;\eta =1;\;k_B=1.$ }
    \label{fig:alpha_dependence_2d}
\end{figure*}

\begin{figure*}
    \centering
    \includegraphics[scale = .6]{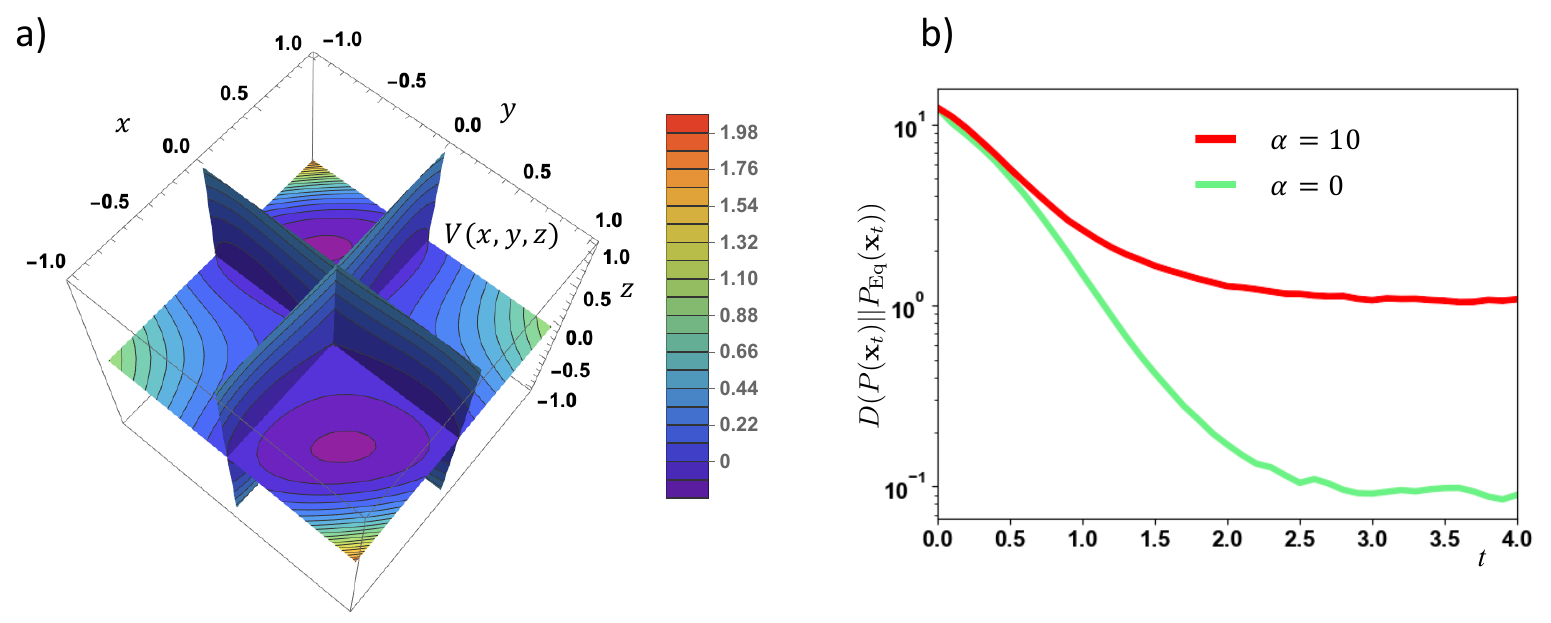}
    \caption{\color{black!0!black} a) The confining potential in Eq.\ \eqref{fig: potential} for $a, \dots, d = 1$ and $\theta = \frac{\pi}{4}$.  b) The total distance function $D \equiv D_{\rm KL}\left( P({\bm x}_t) \vert\vert P_{\rm Eq} ({\bm x}_t) \right)$ for two values of the parameter $\alpha$ ($\alpha = 0$ and $\alpha = 10$). We find that the system with larger value of $\alpha$ takes longer to relax to the stationary state. The distance functions are computed by numerically integrating the Langevin equation in Eq.\ \eqref{eq: 3dl} with time-step $dt = 0.01$, and constructing histograms at different times using $10^5$ copies of trajectories.}
    \label{fig:alpha_dependence_1_3d}
\end{figure*}

\begin{figure*}
    \centering
    \includegraphics[scale = .6]{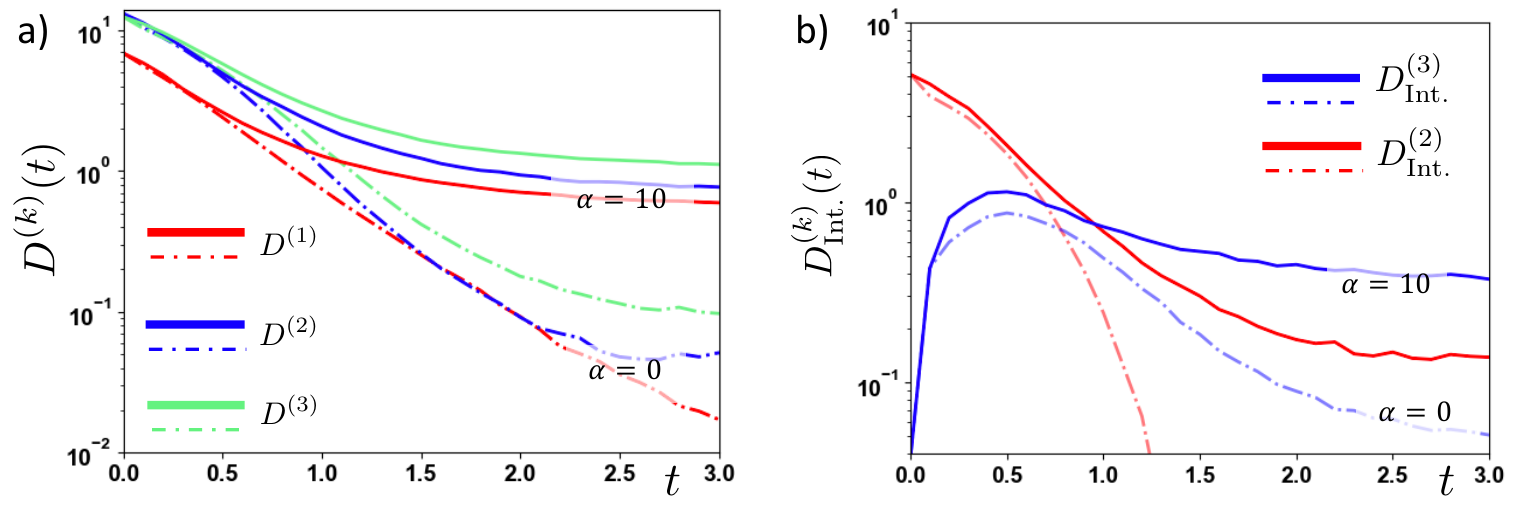}
    \caption{\color{black!0!black} a) The distance functions for different orders of interaction: $D^{(1)} = D_{\rm Ind.}$, $D^{(2)}$, and $D^{(3)} = D$, as well as b) the contributions to total distance arising solely from pairwise and triplet contributions, $D^{(2)}_{\rm Int.}$ and $D^{(3)}_{\rm Int.}$, for $\alpha = 0$ (dot-dashed lines) and $\alpha = 10$ (solid lines). We find that the contributions from interactions, especially triplet interactions, are significantly higher (note the logarithmic scale used for the $y$ axis) when $\alpha = 10$ as compared to the case with $\alpha = 0$. The distance functions are computed by numerically integrating the Langevin equation in Eq.\ \eqref{eq: 3dl} with time-step $dt = 0.01$, and constructing histograms at different times using $10^5$ copies of trajectories.}
    \label{fig:alpha_dependence_2_3d}
\end{figure*}

To demonstrate the usefulness of the decomposition, we {\color{black!0!black} first} consider the problem of thermal relaxation of two identical, {\color{black!0!black}interacting colloidal particles} in two spatially separated quadratic potential wells, as shown in Fig.\ \ref{fig:schematics}. These colloidal particles are prepared in an equilibrium state at temperature $T_0$ and then let to relax in an aqueous solution at temperature $T$. This model has been extensively studied both theoretically \cite{hough2002correlated,kotar2010hydrodynamic,reichert2004hydrodynamic} and experimentally \cite{paul2018two,paul2017direct}. The dynamics is governed by the Langevin equations:\\
\begin{align}
\begin{split}
\label{ eq:l}
 \dot{x}(t)&=H_{11}(-k_1x + f_1(t)) + H_{12}(-k_2y + f_2(t) )\\
 \dot{y}(t)&=H_{21}(-k_1x + f_1(t)) + H_{22}(-k_2y + f_2(t) ),
 \end{split}
\end{align} 
where  $x(t)$ and $y(t)$ are the relative positions of these particles with respect to the center of their respective traps at different times. The parameters $k_1$ and $k_2$ denote the optical stiffness of the two traps. The constants $H_{11} = H_{22} = 1/(6\pi \eta a) = 1/\gamma$ \; and \; $H_{12} = H_{21} = 1/(4\pi \eta R)$, where $R$ is the center - center distance between the two traps and $a$ is the radius of the particle, are the lowest order components, in $1/R$, of the Oseen Tensor \cite{doi1988theory} for motions in the longitudinal directions.  Here $\gamma$ is the viscous drag coefficient. The value of $R$ determines the interaction between the colloidal particles. As $R\to \infty$, the interaction between the colloidal particles vanishes and our system turns to a non-interacting system. The terms $f_1(t)$ and $f_2(t)$ are the random Brownian forces which are delta correlated in time. 

Given that the system is initially prepared in a state different from its thermal equilibrium state in the new environment, it exists in a non-equilibrium state characterized by a certain distance from its eventual thermal state. Quantifying this distance in terms of Kullback-Leibler divergence has gained significant interest in recent times, primarily in the context of non-trivial thermal relaxation behaviours such as Mpemba effects \cite{kumar2020exponentially,bechhoefer2021fresh,biswas2023mpemba,chetrite2021metastable,degunther2022anomalous,ibanez2024heating} or the study of asymmetries of thermal relaxation \cite{Uphill,fewlevelQ,van2021toward,dieball2023asymmetric,meibohm2021relaxation}. In these cases,  $D_{\rm KL}\left( P({\bm x}_t) \vert\vert P_{\rm Eq} ({\bm x}_t) \right)$ is also the same as the excess free energy of the state $P({\bm x}_t)$ \textcolor{black}{which vanishes as the system equilibrates (see Refs.\ \cite{Uphill,chetrite2021metastable} for a simple derivation)}. 

For the model we consider, leveraging the fact that it is a linear system of stochastic differential equations, it is possible to analytically compute the instantaneous probability distribution $P({\bm x}_t)$ in terms of all the parameters in the system, for any value of time $t$ (See Appendix B). 
Using these solutions, it can be verified that the variables $x$ and $y$ are anti-correlated for any $t>0$. The strength of correlations increases when $R$ decreases. Further, at equilibrium (in the $t\rightarrow 0$ and $t\rightarrow \infty$ limit), the correlations vanish.

Using the exact solutions for the distributions, we can further compute  the distance function $D_{\rm KL}\left( P({\bm x}_t) \vert\vert P_{\rm Eq} ({\bm x}_t) \right)$.
In particular, when $t=0$, we get the distance between the initial equilibrium system at temperature $T_0$ and the final equilibrium system at temperature $T$, which can be used to compare initial states and pick the equivalent ones that are equidistant \cite{Uphill,fewlevelQ,van2021toward,meibohm2021relaxation} from the final thermal state. For our model, this initial distance function is found to only depend on the ratio $T_0/T$ and is given by, 
\begin{align}
    D_{\rm KL}\left( P^{T_0}_{\rm Eq}({\bm x}) \vert\vert P_{\rm Eq}^T ({\bm x})\right) = -1+\frac{T_0}{T} +\log \frac{T}{T_0}.
\end{align}
Thus, if we consider an ensemble of systems with different values of $R$, a fixed initial temperature, and an ambient temperature, all of them will have the same distance to the final thermal state at $t=0$. 
For a particular choice of parameters, we show this initial distance function in Fig.\ \ref{fig:schematics}b. The rest of the plots in this paper correspond to the point $T_0/T = 2.5$ in this curve, which has the initial distance $D_{\rm KL}\left( P^{T_0}_{\rm Eq}({\bm x}) \vert\vert P_{\rm Eq}^T ({\bm x})\right) = 0.5837$.

For arbitrary times, the distance functions  $D_{\rm KL}\left( P({\bm x}_t) \vert\vert P_{\rm Eq}^T ({\bm x}_t)\right)$ will in general depend on the the parameter $R$. Furthermore, using explicit analytical solutions of $P({\bm x}_t)$ and its marginals, we can separately compute the independent distance, interaction distance as well as the distance function in the non-interacting limit of $R\rightarrow \infty$. Since our system consists of only two interacting particles,  the decomposition in Eq.\ \eqref{eq:decomposition_new} only has two terms, namely $D^{(1)}_{\rm Int.}=D_{\rm Ind.}$ and $D^{(2)}_{\rm Int.}=D_{\rm Int.} = D - D_{\rm Ind.}$, given by,
\begin{align}
    D_{\rm Int.} = \scalebox{1}{$\int_{x_t,y_t}P(x_t,y_t)\log \frac{P(x_t,y_t)P_{{\rm Eq}, m}(x_t)P_{{\rm Eq},m}(y_t)}{P_{\rm Eq}(x_t,y_t)P_{m}(x_t)P_{m}(y_t)}$}
\end{align}
In Figure \ref{fig:Rt_dependence}, we present our central findings. Figure \ref{fig:Rt_dependence}a illustrates the plots of $D$ and $D_{\rm Ind.}$, for various values of $R$ and $t$, while keeping other model parameters fixed. At $t=0$, all states are equidistant from the final thermal state, as expected. We also find that $D_{\rm Int.} =0$ for any fixed value of $R$.  This means the initial distance function can entirely be determined by the marginal statistics of $x$ and $y$.
However, for $t>0$, and any value of $R$ we observe that $D > D_{\rm Ind.}$, which means interactions positively contribute to the total distance. Specifically, when the two traps are brought closer, the value of $D$ increases for all $t$. Refer to Figure \ref{fig:Rt_dependence}b for a demonstration of this behavior with two different values of $R$.

In Figure \ref{fig:R_dependence}a, we plot the interaction distance $D_{\rm Int.}$ for varying time $t$ and different values of $R$. As $R$ decreases, the interaction distance contribution $D_{\rm Int.}$ increases. Finally, in Figure \ref{fig:R_dependence}b, we compare the total distance $D$ with the distance computed for the non-interacting case, denoted as $D_{\rm non-interacting} \equiv \lim_{R \rightarrow \infty } D $. We observe that $D_{\rm non-interacting} \leq D$ for all values of $R$ and $t$. Moreover, this bound saturates in the limit $R \rightarrow \infty$.

{ \color{black!0!black} So far, we looked at how the interaction parameter $R$ affects the non-equilibrium transformation. It is natural to ask if additional external controls can be introduced in this problem, which affects the rate of transformation, preserving the initial and target states, at a fixed $R$. Interestingly, such a possibility does exist. One can introduce an additional external force of the form ${\bm F}_{\rm ext}({\bm x})= \alpha [ -  \frac{k_2}{\gamma} y,\;  \frac{k_1}{\gamma} x]$, which can be shown to preserve the form of the stationary state  of Eq.\ \eqref{ eq:l} for any fixed $R$, at the cost of making them non-equilibrium with a non-vanishing probability flux and positive entropy production rate \cite{ghimenti2023sampling}. The parameter $\alpha$ can be used to control the strength of this external driving. Once again, the resulting system can be analytically solved and the explicit dependence of the distance functions on the parameter $\alpha$ can be obtained. The results are shown in Fig.\ \ref{fig:alpha_dependence_2d} for a particular choice of $R$ and other system parameters. We find that, as compared to the $\alpha = 0$ case, both the total distance $D$ as well as the interaction distance $D_{\rm Int.}$ is decreased for any value of $t$ as $\alpha$ is increased. This behaviour can further be attributed to the decrease in transient correlations between $x$ and $y$ with increasing $\alpha$ (see Fig.\ \ref{fig:2DCorr} in Appendix B).

{ \color{black!80!black} 
While we have considered a specific form of detailed balance breaking in this example, it's worth noting that for the general class of driven Ornstein-Uhlenbeck processes, as demonstrated in Ref. \cite{Tang}, the non-detailed balance part can always be isolated, regardless of the choice of force and diffusion matrices. This facilitates the construction of a potential function corresponding to the Boltzmann distribution, which remains unaffected by the non-detailed balance contributions. Our formalism can be straightforwardly extended to these cases as well.}

As previously discussed, our general framework extends to interactions beyond second order. To demonstrate this, we now consider a system having three degrees of freedom ${\bm x} = [x, y, z]$ (see also Appendix C), having the following coupled Langevin dynamics:
\begin{align}
\label{eq: 3dl}
    \dot{\bm x} = \left (- {\bm I} + \alpha {\bm Z} \right) \nabla_{\bm x} V(x,y,z) + {\bm \epsilon }(t),
\end{align}
where 
\begin{align}
\label{fig: potential}
\begin{split}
        V(x,y,z) &= a x^{\prime 4} - b x^{\prime 2} + c \frac{y^{\prime 2}}{2} + d\frac{z^2}{2},\\
    \begin{pmatrix}
    x^\prime \\ y^\prime 
    \end{pmatrix}
    &= \begin{pmatrix}
    \cos\theta  &  -\sin\theta\\ \sin\theta & \cos\theta
    \end{pmatrix}
    \times\begin{pmatrix}
    x \\ y
    \end{pmatrix}.
\end{split}
\end{align}
We visualize this potential in Fig.\ \ref{fig:alpha_dependence_1_3d}a. The term ${\bm \epsilon }(t)$ corresponds to Gaussian white noise with $\langle {\bm \epsilon }(t) \rangle = {\bm 0}$ and correlations $\langle {\bm \epsilon}(t){\bm \epsilon}(s) \rangle = 2 {\bm D} \delta(t-s)$. The matrices ${\bm Z}$ and ${\bm D}$ are given by,
\begin{align}
    {\bm Z} &= \begin{bmatrix}
0 & 0 & 1 \\
0  & 0 & 0 \\
 -1 & 0 & 0
\end{bmatrix}, &   {\bm D} &= \left[
\begin{array}{ccc}
 \frac{k_B T}{\gamma } & 0 & 0\\
 0 & \frac{k_B T}{\gamma } & 0\\
 0 & 0 & \frac{k_B T}{  \gamma }
\end{array}
\right].
\end{align}
For arbitrary initial conditions and non-zero values of the constant $\alpha$, and $\theta \in (0, \frac{\pi}{2})$, the system develops both pairwise and triplet correlations. Most of these correlations will be transient, vanishing as the system reaches the stationary state (see Fig.\ \ref{fig:alpha_dependence_corr} in Appendix C. 1). Since the matrix ${\bm Z}$ is skew-symmetric, once again, the stationary state will be non-equilibrium but will have the same form as the Boltzmann-Gibbs distribution with the potential energy function $V$ \cite{ghimenti2023sampling}. Thus, the relaxation process starting from an initial stationary distribution prepared at temperature $T_0$ to a final stationary distribution at temperature $T$, with and without a non-zero $\alpha$, will be {\it equidistant} quenches at $t = 0$.

Due to its non-linearity, tackling this system analytically is challenging. Therefore, we analyze it numerically, and consider the relaxation process corresponding to $\frac{T_0}{T} = 10$. {\color{black!80!black} We provide the corresponding algorithm as a supplementary material \footnote{See Supplemental Material
  at [URL will be inserted by publisher] for a basic implementation of the decomposition of the distance function in python.}.} The results are shown in Figure \ref{fig:alpha_dependence_1_3d}b. We observe that the configuration with external driving ($\alpha = 10$) for which transient correlations develop, takes longer to relax to the stationary state as compared to the configuration without any driving ($\alpha = 0$). We can further use Eq. \eqref{eq:final} through Eq. \eqref{eq:decomposition_new} to compute the distance functions for different orders of interaction: $D^{(1)} = D_{\rm Ind.}$, $D^{(2)}$, and $D^{(3)} = D$, as well as the contributions arising solely from pairwise and triplet contributions, $D^{(2)}_{\rm Int.}$ and $D^{(3)}_{\rm Int.}$. This is demonstrated in Figs.\ \ref{fig:alpha_dependence_2_3d}a and \ref{fig:alpha_dependence_2_3d}b. As expected, we find that the contributions to total distance from interactions, especially triplet interactions, are significantly higher when $\alpha = 10$ as compared to the case with $\alpha = 0$.}

\section{Conclusion}
In summary, we have shown that, in arbitrary non-equilibrium transformations, the distance between the current state and a target state can be decomposed into two terms: one corresponding to an \textit{independent} estimate of the distance, representing hypothetical marginal processes which are non-interacting, and another corresponding to interactions, quantified using the relative mutual information between the variables. The interaction term can further be decomposed into contributions from interactions between pairs of elements, interactions among triplets, and so on. The results are demonstrated by considering, {\color{black!0!black} \textit{a)} the example of the thermal relaxation of two interacting optically trapped colloidal particles, and \textit{b)} a three dimensional system driven by non-conservative forces. In both cases, it is observed that increasing the interaction strength enhances transient correlations, increasing the separation between the time-dependent non-equilibrium state and the target state. Moreover, for fixed values of interaction parameters, our formalism separates out the contributions to the total distance, at any time, arising from different orders of interactions between the variables. The results also show that introducing additional non-conservative driving forces provides an extra degree of control over the transformation process.}

Our results suggest that harnessing local interactions could have applications in controlling and taming the time evolution of systems towards desired states.  In setups where it is possible to control the strength of different orders of interactions, our findings offer a possible way to disentangle their effects on the transformation process, and to identify the ones that can assist the transformation. As mentioned, our decomposition of the distance function is not necessarily unique but merits further investigation in interacting systems with many degrees of freedom. Further research could also delve into specific applications in non-equilibrium control problems \cite{aurell2011optimal,yan2022learning,rotskoff2017geometric,abreu2012thermodynamics,chennakesavalu2023unified} where understanding these effects could be valuable, or resource theories \cite{gour2015resource}, where maintaining non-equilibrium states for extended periods could be beneficial.

\section*{Acknowledgements}
MR and SKM thank the Kerala Theoretical Physics Initiative - Active Research Training (KTPI - ART) program for facilitating the research collaboration. 
SKM acknowledges the Knut and Alice Wallenberg Foundation for financial support through Grant No. KAW 2021.0328. SKM thanks the members of the Soft-Matter Group, NORDITA, Stockholm, Sweden, for helpful discussions on Refs. \cite{lynn2022emergence,arrowoftime}. SKM thanks Biswajit Das and Shuvojit Paul, Light Matter Lab, IISER Kolkata, India for helpful discussions on the model studied. SKM thanks Clay Batton for feedback on an earlier version of this manuscript.

\appendix
{\color{black!0!black}

\section{Kirkwood approximation upto order $k=3$}
Here we provide the form of Eq.\ \eqref{eq:final} for $k = 1, 2$ and $3$ (see the main text for notations). Let $P({\bm x})$ be a joint distributions of $N$ variables. The corresponding $k = 1$ approximation is just the product of the single variable marginals, given as,
\begin{align}
  \log  \hat{P}_1 ({\bm x}) = \log \left[ \prod_i^N P_1(x_i) \right]
\end{align}
The $k = 2$ approximation is:
\begin{align}
  \log  \hat{P}_2 ({\bm x}) = \log \left[ \prod_{i>j}  \frac{P_2(x_i,x_j)}{P_1(x_i)} \right]
\end{align}
The $k = 3$ approximation is:
\begin{align}
  \log  \hat{P}_3({\bm x}) = \scalebox{1}{$ \log \left[ \prod_{i>j>k}  \frac{P_3(x_i,x_j,x_k)P_1(x_k)P_1(x_j)}{P_2(x_i,x_k)P_2(x_j,x_k)} \right]$}
\end{align}
Higher order approximations can be similarly obtained by applying Eq.\ \eqref{eq:final}.}
\section{Exact calculation for the system of interacting colloids}
Here, we describe the calculation of the distance functions for the model of interacting colloids. We follow the notations in Ref.\ \cite{aykut}. To begin with, we rewrite Eq.\ \eqref{ eq:l} as a matrix equation,
\begin{equation}
    \dot {{\bm r}}(t) = - {\bm A} {\bm r}(t) + {\bm \epsilon}(t)
\end{equation}\\
with $\langle {\bm \epsilon}(t){\bm \epsilon}(s) \rangle = 2 {\bm D} \delta(t-s)$ where 
\begin{align}
\begin{split}
{\bm r}(t)&= \begin{bmatrix}
x(t) \\
y(t)
\end{bmatrix},\\ {\bm A} &= \begin{bmatrix}
\frac{k_1}{\gamma} & \frac{k_2}{4\pi \eta R} \\
\frac{k_1}{4\pi \eta R}  & \frac{k_2}{\gamma}
\end{bmatrix},\\ {\bm D} &= \left[
\begin{array}{cc}
 \frac{k_B T}{\gamma } & \frac{k_B T}{4 \pi  \eta  R} \\
 \frac{k_B T}{4 \pi  \eta  R} & \frac{k_B T}{\gamma } \\
\end{array}
\right], \\
{\bm \eta}(t)&= \begin{bmatrix}
\epsilon_1(t) \\
\epsilon_2(t)
\end{bmatrix}.
\end{split}
\end{align}
{\color{black!0!black} For the case with non-equilibrium driving, we consider the case where 
\begin{align}
    {\bm A} &= \begin{bmatrix}
\frac{k_1}{\gamma} & \frac{k_2}{4\pi \eta R} + \alpha \; \frac{k_2}{\gamma} \\
\frac{k_1}{4\pi \eta R} -\alpha \; \frac{k_1}{\gamma}  & \frac{k_2}{\gamma}
\end{bmatrix}
\end{align}
}
Now to find the probability distribution of the system at any time, first, we write a Fokker - Planck equation equivalent to our Langevin equation as \\
\begin{equation}\label{A.4}
    \frac{\partial P(t,{\bm r}\vert t_0,{\bm r}_0)}{\partial t} = \sum_{i,j}(\frac{\partial}{\partial r_i}[A_{ij} r_j + D_{ij} \frac{\partial}{\partial r_j}]P(t,{\bm r}\vert t_0,{\bm r}_0)) \;,\\ 
\end{equation}
where $P(t,{\bm r}|t_0,{\bm r}_0)$ is  the conditional probability that the system is in a position ${\bm r}$ at time $t$, given that it was at ${\bm r}_0$ at time $t_0$. 

The Fokker - Planck equation (\ref{A.4}) is exactly solvable, and the solution is found to be \\
\begin{equation}\label{eq:transitionProb}
 P(t,{\bm r}|t_0,{\bm r}_0)  = \scalebox{0.9}{$\frac{e^{-\frac{1}{2} [ {\bm r} - e^{-(t-t_0){\bm A}}{\bm r}_0]^T\; {\bm \Sigma}^{-1}(t-t_0) \;[{\bm r} - e^{-(t-t_0){\bm A}}{\bm r}_0]}}{\sqrt{(2 \pi )^2\; \det \;{\bm \Sigma}(t-t_0)}}$} \; ,
\end{equation}\\
where the covariance matrix is \\
\begin{equation}
    {\bm \Sigma}(t)= {\bm \Sigma} (\infty) - e^{-t {\bm A}}\; {\bm \Sigma}(\infty)\; e^{-t {\bm A}^T} \;, 
\end{equation}
and ${\bm \Sigma}(\infty)$ is found by solving the below matrix equation
\begin{equation}
    {\bm A} {\bm \Sigma}(\infty) + {\bm \Sigma}(\infty) {\bm A}^T = 2{\bm D}\\
\end{equation}
If the matrix ${\bm A}$ is positive definite, it is guaranteed that the system will reach a stationary Gaussian distribution at $t \rightarrow \infty$, which will have the covariance matrix ${\bm \Sigma}^{-1}(\infty)$.
For our model, we obtain,
\begin{align}
    {\bm \Sigma}(\infty) =\left(
\begin{array}{cc}
 \frac{k_B T }{k_1} & 0 \\
 0 & \frac{k_B T }{k_2} 
\end{array}
\right).
\end{align}
In terms of this matrix, we can obtain the equilibrium distribution of the system as, 
\begin{align}
    P_{\rm Eq}({\bm x})= \frac{1}{\sqrt{(2\pi)^2 \det {\bm \Sigma}(\infty)}} e^{-\frac{1 }{2} {\bm x} {\bm \Sigma}^{-1}(\infty) {\bm x}}.
\end{align}
Note that this distribution explicitly depends on the temperature $T$. When we set $T=T_0$, we get the equilibrium distribution at temperature $T_0$. Furthermore, the time-dependent distribution corresponding to the thermal relaxation from a distribution at an initial temperature $T_0$ to an ambient temperature $T$ can be obtained by performing the integration, 
\begin{align}
    P({\bm x}_t) = \int_{{\bm x}_0} P_{\rm Eq}^{T_0}({\bm x}_0) P(t,{\bm x}_t\vert t_0,{\bm x}_0)
\end{align}
where $P(t,{\bm x}_t\vert t_0,{\bm x}_0)$ is given by Eq.\ \eqref{eq:transitionProb}. The results in this manuscript are obtained by first explicitly evaluating this integral to get $P({\bm x}_t)$ and computing the relevant distance functions in terms of that. 
{ \color{black!0!black}
\subsection*{B. 1: The correlations between $x$ and $y$}
\label{2dcorr}
Figure \ref{fig:2DCorr} shows the correlations between $x$ and $y$ variables as a function of $t$ for a fixed $R$ and varying values of $\alpha$.
\begin{figure}
    \centering
    \includegraphics[width = 0.8\linewidth]{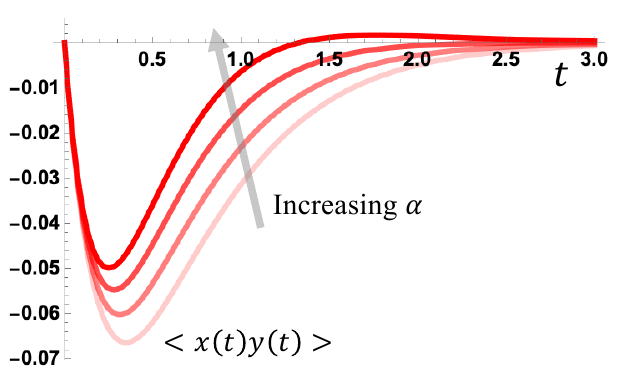}
    \caption{\color{black!0!black} Figure showing the dependence of the $x$, $y$ correlations on $\alpha$ as a function of time $t$ for a fixed value of $R$. The other parameter choices are: $k_1=1;\;k_2=2;\;\gamma =1;\;\eta =1;\;k_B=1.$ The values of $\alpha$ considered are $\alpha = 0,0.05,0.1, 0.15$. }
    \label{fig:2DCorr}
\end{figure}
\section{Example of a system with pairwise and triplet interactions}
\label{3dint}
As an example of a system with third order interactions, we consider a system with three degrees of freedom $(x,\;y,\;z)$ having a confining potential:
\begin{align}
\begin{split}
        V(x,y,z) &= a x^{\prime 4} - b x^{\prime 2} + c \frac{y^{\prime 2}}{2} + d\frac{z^2}{2},\\
    \begin{pmatrix}
    x^\prime \\ y^\prime 
    \end{pmatrix}
    &= \begin{pmatrix}
    \cos\theta  &  -\sin\theta\\ \sin\theta & \cos\theta
    \end{pmatrix}
    \times\begin{pmatrix}
    x \\ y
    \end{pmatrix},
\end{split}
\end{align}
where the rotation matrix is used to couple the $x$ and $y$ degrees of freedom. The parameters $a,\dots, d$ needs to be chosen such that the overall potential is confining. We set all these parameters to $1$ such that the confinement along the $x^\prime$ direction corresponds to a double-well potential. Next, we consider the overdamped Langevin dynamics of this system:
\begin{align}
    \dot{\bm x} = \left (- {\bm I} + \alpha {\bm Z} \right) \nabla_{\bm x} V(x,y,z) + {\bm \epsilon }(t),
\end{align}
where ${\bm I}$ is the identity matrix and ${\bm Z}$ is any skew-symmetric matrix which will lead to a non-conservative driving. Interestingly, it can be shown that this additional driving does not change the stationary state of the system from the Boltzmann distribution \cite{ghimenti2023sampling}. Here, the parameter $\alpha$ determines the strength of this driving. The noise correlations are given by $\langle {\bm \epsilon}(t){\bm \epsilon}(s) \rangle = 2 {\bm D} \delta(t-s)$ where,
\begin{align}
    {\bm D} &= \left[
\begin{array}{ccc}
 \frac{k_B T}{\gamma } & 0 & 0\\
 0 & \frac{k_B T}{\gamma } & 0\\
 0 & 0 & \frac{k_B T}{  \gamma }
\end{array}
\right].
\end{align}
In particular, we choose
\begin{align}
    {\bm Z} &= \begin{bmatrix}
0 & 0 & 1 \\
0  & 0 & 0 \\
 -1 & 0 & 0
\end{bmatrix},
\end{align}
which effectively couples the $x, y$ degrees to $z$, leading to new pair-wise and triplet interactions. A scenario where the effects of the interactions can be seen is when you consider a thermal relaxation dynamics, where
we prepare the system at an arbitrary initial temperature $T_0$ and let it relax to the steady state at an ambient temperature $T$. In addition, if we choose a non-zero $\alpha$, transient correlations develop between all the three variables. Similar to the two-particle case, the choice of $\alpha$ does not affect initial and final distributions, ensuring that the distance function at $t=0$ remains independent of $\alpha$. This facilitates the comparison of initially equivalent states. 

\subsection*{C. 1: The correlations between $x$, $y$ and $z$}
\label{3dcorr}
Figure \ref{fig:alpha_dependence_corr} shows various correlations between $x$, $y$ and $z$ variables as a function of $t$ for two different values of $\alpha$ and other parameters fixed.
\begin{figure*}
    \centering
    \includegraphics[width = 0.8\linewidth]{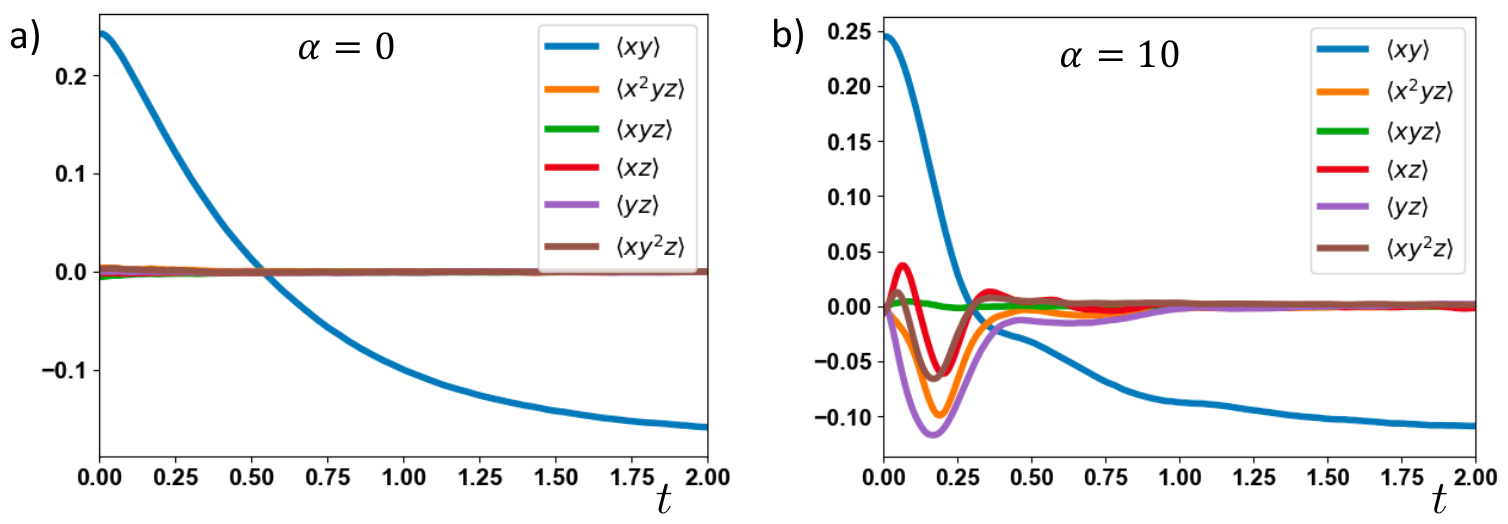}
    \caption{ \color{black!0!black} The figure illustrates various correlations of the $x$, $y$, and $z$ variables for the dynamical system in Eq.\ \eqref{eq: 3dl} as a function of time, $t$, for two different values of $\alpha$: a) $\alpha =0$ and b) $\alpha = 10$. The other parameters are kept fixed ($a=1;\;b=1;\;c = 1,\; \theta = \frac{\pi}{4},\; d = 1,\; \gamma =1;\;\eta =1;\;k_B=1, \; T_0 = 1,\; T = \frac{1}{10}$). The correlation functions are computed by numerically integrating the Langevin equation in Eq.\ \eqref{eq: 3dl} with time-step $dt = 0.01$, and  using $10^5$ copies of trajectories.}
    \label{fig:alpha_dependence_corr}
\end{figure*}

}


\begin{thebibliography}{69}%
\makeatletter
\providecommand \@ifxundefined [1]{%
 \@ifx{#1\undefined}
}%
\providecommand \@ifnum [1]{%
 \ifnum #1\expandafter \@firstoftwo
 \else \expandafter \@secondoftwo
 \fi
}%
\providecommand \@ifx [1]{%
 \ifx #1\expandafter \@firstoftwo
 \else \expandafter \@secondoftwo
 \fi
}%
\providecommand \natexlab [1]{#1}%
\providecommand \enquote  [1]{``#1''}%
\providecommand \bibnamefont  [1]{#1}%
\providecommand \bibfnamefont [1]{#1}%
\providecommand \citenamefont [1]{#1}%
\providecommand \href@noop [0]{\@secondoftwo}%
\providecommand \href [0]{\begingroup \@sanitize@url \@href}%
\providecommand \@href[1]{\@@startlink{#1}\@@href}%
\providecommand \@@href[1]{\endgroup#1\@@endlink}%
\providecommand \@sanitize@url [0]{\catcode `\\12\catcode `\$12\catcode
  `\&12\catcode `\#12\catcode `\^12\catcode `\_12\catcode `\%12\relax}%
\providecommand \@@startlink[1]{}%
\providecommand \@@endlink[0]{}%
\providecommand \url  [0]{\begingroup\@sanitize@url \@url }%
\providecommand \@url [1]{\endgroup\@href {#1}{\urlprefix }}%
\providecommand \urlprefix  [0]{URL }%
\providecommand \Eprint [0]{\href }%
\providecommand \doibase [0]{https://doi.org/}%
\providecommand \selectlanguage [0]{\@gobble}%
\providecommand \bibinfo  [0]{\@secondoftwo}%
\providecommand \bibfield  [0]{\@secondoftwo}%
\providecommand \translation [1]{[#1]}%
\providecommand \BibitemOpen [0]{}%
\providecommand \bibitemStop [0]{}%
\providecommand \bibitemNoStop [0]{.\EOS\space}%
\providecommand \EOS [0]{\spacefactor3000\relax}%
\providecommand \BibitemShut  [1]{\csname bibitem#1\endcsname}%
\let\auto@bib@innerbib\@empty
\bibitem [{\citenamefont {Dattagupta}(2012)}]{dattagupta2012relaxation}%
  \BibitemOpen
  \bibfield  {author} {\bibinfo {author} {\bibfnamefont {S.}~\bibnamefont
  {Dattagupta}},\ }\href@noop {} {\emph {\bibinfo {title} {Relaxation phenomena
  in condensed matter physics}}}\ (\bibinfo  {publisher} {Elsevier},\ \bibinfo
  {year} {2012})\BibitemShut {NoStop}%
\bibitem [{\citenamefont {Whitesides}\ and\ \citenamefont
  {Grzybowski}(2002)}]{whitesides2002self}%
  \BibitemOpen
  \bibfield  {author} {\bibinfo {author} {\bibfnamefont {G.~M.}\ \bibnamefont
  {Whitesides}}\ and\ \bibinfo {author} {\bibfnamefont {B.}~\bibnamefont
  {Grzybowski}},\ }\bibfield  {title} {\bibinfo {title} {Self-assembly at all
  scales},\ }\href@noop {} {\bibfield  {journal} {\bibinfo  {journal}
  {Science}\ }\textbf {\bibinfo {volume} {295}},\ \bibinfo {pages} {2418}
  (\bibinfo {year} {2002})}\BibitemShut {NoStop}%
\bibitem [{\citenamefont {Pollard}\ and\ \citenamefont
  {Borisy}(2003)}]{pollard_cellular_2003}%
  \BibitemOpen
  \bibfield  {author} {\bibinfo {author} {\bibfnamefont {T.~D.}\ \bibnamefont
  {Pollard}}\ and\ \bibinfo {author} {\bibfnamefont {G.~G.}\ \bibnamefont
  {Borisy}},\ }\bibfield  {title} {\bibinfo {title} {Cellular {{Motility
  Driven}} by {{Assembly}} and {{Disassembly}} of {{Actin Filaments}}},\ }\href
  {https://doi.org/10.1016/S0092-8674(03)00120-X} {\bibfield  {journal}
  {\bibinfo  {journal} {Cell}\ }\textbf {\bibinfo {volume} {112}},\ \bibinfo
  {pages} {453} (\bibinfo {year} {2003})}\BibitemShut {NoStop}%
\bibitem [{\citenamefont {Mauro}\ \emph {et~al.}(2014)\citenamefont {Mauro},
  \citenamefont {Aliprandi}, \citenamefont {Septiadi}, \citenamefont {Kehr},\
  and\ \citenamefont {De~Cola}}]{mauro2014self}%
  \BibitemOpen
  \bibfield  {author} {\bibinfo {author} {\bibfnamefont {M.}~\bibnamefont
  {Mauro}}, \bibinfo {author} {\bibfnamefont {A.}~\bibnamefont {Aliprandi}},
  \bibinfo {author} {\bibfnamefont {D.}~\bibnamefont {Septiadi}}, \bibinfo
  {author} {\bibfnamefont {N.~S.}\ \bibnamefont {Kehr}},\ and\ \bibinfo
  {author} {\bibfnamefont {L.}~\bibnamefont {De~Cola}},\ }\bibfield  {title}
  {\bibinfo {title} {When self-assembly meets biology: luminescent platinum
  complexes for imaging applications},\ }\href@noop {} {\bibfield  {journal}
  {\bibinfo  {journal} {Chemical Society Reviews}\ }\textbf {\bibinfo {volume}
  {43}},\ \bibinfo {pages} {4144} (\bibinfo {year} {2014})}\BibitemShut
  {NoStop}%
\bibitem [{\citenamefont {Dobson}(2003)}]{dobson2003protein}%
  \BibitemOpen
  \bibfield  {author} {\bibinfo {author} {\bibfnamefont {C.~M.}\ \bibnamefont
  {Dobson}},\ }\bibfield  {title} {\bibinfo {title} {Protein folding and
  misfolding},\ }\href@noop {} {\bibfield  {journal} {\bibinfo  {journal}
  {Nature}\ }\textbf {\bibinfo {volume} {426}},\ \bibinfo {pages} {884}
  (\bibinfo {year} {2003})}\BibitemShut {NoStop}%
\bibitem [{\citenamefont {Creighton}(1990)}]{creighton1990protein}%
  \BibitemOpen
  \bibfield  {author} {\bibinfo {author} {\bibfnamefont {T.~E.}\ \bibnamefont
  {Creighton}},\ }\bibfield  {title} {\bibinfo {title} {Protein folding.},\
  }\href@noop {} {\bibfield  {journal} {\bibinfo  {journal} {Biochemical
  journal}\ }\textbf {\bibinfo {volume} {270}},\ \bibinfo {pages} {1} (\bibinfo
  {year} {1990})}\BibitemShut {NoStop}%
\bibitem [{\citenamefont {Ritort}(2006)}]{ritort2006single}%
  \BibitemOpen
  \bibfield  {author} {\bibinfo {author} {\bibfnamefont {F.}~\bibnamefont
  {Ritort}},\ }\bibfield  {title} {\bibinfo {title} {Single-molecule
  experiments in biological physics: methods and applications},\ }\href@noop {}
  {\bibfield  {journal} {\bibinfo  {journal} {Journal of Physics: Condensed
  Matter}\ }\textbf {\bibinfo {volume} {18}},\ \bibinfo {pages} {R531}
  (\bibinfo {year} {2006})}\BibitemShut {NoStop}%
\bibitem [{\citenamefont {Ciliberto}(2017)}]{ciliberto2017experiments}%
  \BibitemOpen
  \bibfield  {author} {\bibinfo {author} {\bibfnamefont {S.}~\bibnamefont
  {Ciliberto}},\ }\bibfield  {title} {\bibinfo {title} {Experiments in
  stochastic thermodynamics: Short history and perspectives},\ }\href@noop {}
  {\bibfield  {journal} {\bibinfo  {journal} {Physical Review X}\ }\textbf
  {\bibinfo {volume} {7}},\ \bibinfo {pages} {021051} (\bibinfo {year}
  {2017})}\BibitemShut {NoStop}%
\bibitem [{\citenamefont {Pop}(2010)}]{pop2010energy}%
  \BibitemOpen
  \bibfield  {author} {\bibinfo {author} {\bibfnamefont {E.}~\bibnamefont
  {Pop}},\ }\bibfield  {title} {\bibinfo {title} {Energy dissipation and
  transport in nanoscale devices},\ }\href@noop {} {\bibfield  {journal}
  {\bibinfo  {journal} {Nano Research}\ }\textbf {\bibinfo {volume} {3}},\
  \bibinfo {pages} {147} (\bibinfo {year} {2010})}\BibitemShut {NoStop}%
\bibitem [{\citenamefont {Bergfield}\ and\ \citenamefont
  {Ratner}(2013)}]{bergfield2013forty}%
  \BibitemOpen
  \bibfield  {author} {\bibinfo {author} {\bibfnamefont {J.~P.}\ \bibnamefont
  {Bergfield}}\ and\ \bibinfo {author} {\bibfnamefont {M.~A.}\ \bibnamefont
  {Ratner}},\ }\bibfield  {title} {\bibinfo {title} {Forty years of molecular
  electronics: Non-equilibrium heat and charge transport at the nanoscale},\
  }\href@noop {} {\bibfield  {journal} {\bibinfo  {journal} {physica status
  solidi (b)}\ }\textbf {\bibinfo {volume} {250}},\ \bibinfo {pages} {2249}
  (\bibinfo {year} {2013})}\BibitemShut {NoStop}%
\bibitem [{\citenamefont {Mart{\'\i}nez}\ \emph {et~al.}(2016)\citenamefont
  {Mart{\'\i}nez}, \citenamefont {Rold{\'a}n}, \citenamefont {Dinis},
  \citenamefont {Petrov}, \citenamefont {Parrondo},\ and\ \citenamefont
  {Rica}}]{martinez2016brownian}%
  \BibitemOpen
  \bibfield  {author} {\bibinfo {author} {\bibfnamefont {I.~A.}\ \bibnamefont
  {Mart{\'\i}nez}}, \bibinfo {author} {\bibfnamefont {{\'E}.}~\bibnamefont
  {Rold{\'a}n}}, \bibinfo {author} {\bibfnamefont {L.}~\bibnamefont {Dinis}},
  \bibinfo {author} {\bibfnamefont {D.}~\bibnamefont {Petrov}}, \bibinfo
  {author} {\bibfnamefont {J.~M.}\ \bibnamefont {Parrondo}},\ and\ \bibinfo
  {author} {\bibfnamefont {R.~A.}\ \bibnamefont {Rica}},\ }\bibfield  {title}
  {\bibinfo {title} {Brownian carnot engine},\ }\href@noop {} {\bibfield
  {journal} {\bibinfo  {journal} {Nature physics}\ }\textbf {\bibinfo {volume}
  {12}},\ \bibinfo {pages} {67} (\bibinfo {year} {2016})}\BibitemShut {NoStop}%
\bibitem [{\citenamefont {Sagawa}\ and\ \citenamefont
  {Ueda}(2012)}]{sagawa2012fluctuation}%
  \BibitemOpen
  \bibfield  {author} {\bibinfo {author} {\bibfnamefont {T.}~\bibnamefont
  {Sagawa}}\ and\ \bibinfo {author} {\bibfnamefont {M.}~\bibnamefont {Ueda}},\
  }\bibfield  {title} {\bibinfo {title} {Fluctuation theorem with information
  exchange: Role of correlations in stochastic thermodynamics},\ }\href@noop {}
  {\bibfield  {journal} {\bibinfo  {journal} {Physical review letters}\
  }\textbf {\bibinfo {volume} {109}},\ \bibinfo {pages} {180602} (\bibinfo
  {year} {2012})}\BibitemShut {NoStop}%
\bibitem [{\citenamefont {Barato}\ and\ \citenamefont
  {Seifert}(2014)}]{barato2014unifying}%
  \BibitemOpen
  \bibfield  {author} {\bibinfo {author} {\bibfnamefont {A.}~\bibnamefont
  {Barato}}\ and\ \bibinfo {author} {\bibfnamefont {U.}~\bibnamefont
  {Seifert}},\ }\bibfield  {title} {\bibinfo {title} {Unifying three
  perspectives on information processing in stochastic thermodynamics},\
  }\href@noop {} {\bibfield  {journal} {\bibinfo  {journal} {Physical review
  letters}\ }\textbf {\bibinfo {volume} {112}},\ \bibinfo {pages} {090601}
  (\bibinfo {year} {2014})}\BibitemShut {NoStop}%
\bibitem [{\citenamefont {Aurell}\ \emph {et~al.}(2012)\citenamefont {Aurell},
  \citenamefont {Gawedzki}, \citenamefont {Mejia-Monasterio}, \citenamefont
  {Mohayaee},\ and\ \citenamefont {Muratore-Ginanneschi}}]{aurell2012refined}%
  \BibitemOpen
  \bibfield  {author} {\bibinfo {author} {\bibfnamefont {E.}~\bibnamefont
  {Aurell}}, \bibinfo {author} {\bibfnamefont {K.}~\bibnamefont {Gawedzki}},
  \bibinfo {author} {\bibfnamefont {C.}~\bibnamefont {Mejia-Monasterio}},
  \bibinfo {author} {\bibfnamefont {R.}~\bibnamefont {Mohayaee}},\ and\
  \bibinfo {author} {\bibfnamefont {P.}~\bibnamefont {Muratore-Ginanneschi}},\
  }\bibfield  {title} {\bibinfo {title} {Refined second law of thermodynamics
  for fast random processes},\ }\href@noop {} {\bibfield  {journal} {\bibinfo
  {journal} {Journal of statistical physics}\ }\textbf {\bibinfo {volume}
  {147}},\ \bibinfo {pages} {487} (\bibinfo {year} {2012})}\BibitemShut
  {NoStop}%
\bibitem [{\citenamefont {Deffner}\ and\ \citenamefont {Lutz}(2010)}]{Lutz}%
  \BibitemOpen
  \bibfield  {author} {\bibinfo {author} {\bibfnamefont {S.}~\bibnamefont
  {Deffner}}\ and\ \bibinfo {author} {\bibfnamefont {E.}~\bibnamefont {Lutz}},\
  }\bibfield  {title} {\bibinfo {title} {Generalized clausius inequality for
  nonequilibrium quantum processes},\ }\href
  {https://doi.org/10.1103/PhysRevLett.105.170402} {\bibfield  {journal}
  {\bibinfo  {journal} {Phys. Rev. Lett.}\ }\textbf {\bibinfo {volume} {105}},\
  \bibinfo {pages} {170402} (\bibinfo {year} {2010})}\BibitemShut {NoStop}%
\bibitem [{\citenamefont {Kim}(2021)}]{kim2021information}%
  \BibitemOpen
  \bibfield  {author} {\bibinfo {author} {\bibfnamefont {E.-j.}\ \bibnamefont
  {Kim}},\ }\bibfield  {title} {\bibinfo {title} {Information geometry,
  fluctuations, non-equilibrium thermodynamics, and geodesics in complex
  systems},\ }\href@noop {} {\bibfield  {journal} {\bibinfo  {journal}
  {Entropy}\ }\textbf {\bibinfo {volume} {23}},\ \bibinfo {pages} {1393}
  (\bibinfo {year} {2021})}\BibitemShut {NoStop}%
\bibitem [{\citenamefont {Nakazato}\ and\ \citenamefont
  {Ito}(2021)}]{nakazato2021geometrical}%
  \BibitemOpen
  \bibfield  {author} {\bibinfo {author} {\bibfnamefont {M.}~\bibnamefont
  {Nakazato}}\ and\ \bibinfo {author} {\bibfnamefont {S.}~\bibnamefont {Ito}},\
  }\bibfield  {title} {\bibinfo {title} {Geometrical aspects of entropy
  production in stochastic thermodynamics based on wasserstein distance},\
  }\href@noop {} {\bibfield  {journal} {\bibinfo  {journal} {Physical Review
  Research}\ }\textbf {\bibinfo {volume} {3}},\ \bibinfo {pages} {043093}
  (\bibinfo {year} {2021})}\BibitemShut {NoStop}%
\bibitem [{\citenamefont {Ito}(2023)}]{ito2023geometric}%
  \BibitemOpen
  \bibfield  {author} {\bibinfo {author} {\bibfnamefont {S.}~\bibnamefont
  {Ito}},\ }\bibfield  {title} {\bibinfo {title} {Geometric thermodynamics for
  the fokker--planck equation: stochastic thermodynamic links between
  information geometry and optimal transport},\ }\href@noop {} {\bibfield
  {journal} {\bibinfo  {journal} {Information Geometry}\ ,\ \bibinfo {pages}
  {1}} (\bibinfo {year} {2023})}\BibitemShut {NoStop}%
\bibitem [{\citenamefont {Chennakesavalu}\ and\ \citenamefont
  {Rotskoff}(2023)}]{chennakesavalu2023unified}%
  \BibitemOpen
  \bibfield  {author} {\bibinfo {author} {\bibfnamefont {S.}~\bibnamefont
  {Chennakesavalu}}\ and\ \bibinfo {author} {\bibfnamefont {G.~M.}\
  \bibnamefont {Rotskoff}},\ }\bibfield  {title} {\bibinfo {title} {Unified,
  geometric framework for nonequilibrium protocol optimization},\ }\href@noop
  {} {\bibfield  {journal} {\bibinfo  {journal} {Physical Review Letters}\
  }\textbf {\bibinfo {volume} {130}},\ \bibinfo {pages} {107101} (\bibinfo
  {year} {2023})}\BibitemShut {NoStop}%
\bibitem [{\citenamefont {Rotskoff}\ and\ \citenamefont
  {Crooks}(2015)}]{rotskoff2015optimal}%
  \BibitemOpen
  \bibfield  {author} {\bibinfo {author} {\bibfnamefont {G.~M.}\ \bibnamefont
  {Rotskoff}}\ and\ \bibinfo {author} {\bibfnamefont {G.~E.}\ \bibnamefont
  {Crooks}},\ }\bibfield  {title} {\bibinfo {title} {Optimal control in
  nonequilibrium systems: Dynamic riemannian geometry of the ising model},\
  }\href@noop {} {\bibfield  {journal} {\bibinfo  {journal} {Physical Review
  E}\ }\textbf {\bibinfo {volume} {92}},\ \bibinfo {pages} {060102} (\bibinfo
  {year} {2015})}\BibitemShut {NoStop}%
\bibitem [{\citenamefont {Shiraishi}\ \emph {et~al.}(2018)\citenamefont
  {Shiraishi}, \citenamefont {Funo},\ and\ \citenamefont
  {Saito}}]{shiraishi2018speed}%
  \BibitemOpen
  \bibfield  {author} {\bibinfo {author} {\bibfnamefont {N.}~\bibnamefont
  {Shiraishi}}, \bibinfo {author} {\bibfnamefont {K.}~\bibnamefont {Funo}},\
  and\ \bibinfo {author} {\bibfnamefont {K.}~\bibnamefont {Saito}},\ }\bibfield
   {title} {\bibinfo {title} {Speed limit for classical stochastic processes},\
  }\href@noop {} {\bibfield  {journal} {\bibinfo  {journal} {Physical review
  letters}\ }\textbf {\bibinfo {volume} {121}},\ \bibinfo {pages} {070601}
  (\bibinfo {year} {2018})}\BibitemShut {NoStop}%
\bibitem [{\citenamefont {Van~Vu}\ and\ \citenamefont
  {Saito}(2023{\natexlab{a}})}]{van2023topological}%
  \BibitemOpen
  \bibfield  {author} {\bibinfo {author} {\bibfnamefont {T.}~\bibnamefont
  {Van~Vu}}\ and\ \bibinfo {author} {\bibfnamefont {K.}~\bibnamefont {Saito}},\
  }\bibfield  {title} {\bibinfo {title} {Topological speed limit},\ }\href@noop
  {} {\bibfield  {journal} {\bibinfo  {journal} {Physical review letters}\
  }\textbf {\bibinfo {volume} {130}},\ \bibinfo {pages} {010402} (\bibinfo
  {year} {2023}{\natexlab{a}})}\BibitemShut {NoStop}%
\bibitem [{\citenamefont {Yoshimura}\ and\ \citenamefont
  {Ito}(2021)}]{yoshimura2021thermodynamic}%
  \BibitemOpen
  \bibfield  {author} {\bibinfo {author} {\bibfnamefont {K.}~\bibnamefont
  {Yoshimura}}\ and\ \bibinfo {author} {\bibfnamefont {S.}~\bibnamefont
  {Ito}},\ }\bibfield  {title} {\bibinfo {title} {Thermodynamic uncertainty
  relation and thermodynamic speed limit in deterministic chemical reaction
  networks},\ }\href@noop {} {\bibfield  {journal} {\bibinfo  {journal}
  {Physical review letters}\ }\textbf {\bibinfo {volume} {127}},\ \bibinfo
  {pages} {160601} (\bibinfo {year} {2021})}\BibitemShut {NoStop}%
\bibitem [{\citenamefont {Funo}\ \emph {et~al.}(2019)\citenamefont {Funo},
  \citenamefont {Shiraishi},\ and\ \citenamefont {Saito}}]{funo2019speed}%
  \BibitemOpen
  \bibfield  {author} {\bibinfo {author} {\bibfnamefont {K.}~\bibnamefont
  {Funo}}, \bibinfo {author} {\bibfnamefont {N.}~\bibnamefont {Shiraishi}},\
  and\ \bibinfo {author} {\bibfnamefont {K.}~\bibnamefont {Saito}},\ }\bibfield
   {title} {\bibinfo {title} {Speed limit for open quantum systems},\
  }\href@noop {} {\bibfield  {journal} {\bibinfo  {journal} {New Journal of
  Physics}\ }\textbf {\bibinfo {volume} {21}},\ \bibinfo {pages} {013006}
  (\bibinfo {year} {2019})}\BibitemShut {NoStop}%
\bibitem [{\citenamefont {Lee}\ \emph {et~al.}(2022)\citenamefont {Lee},
  \citenamefont {Lee}, \citenamefont {Kwon},\ and\ \citenamefont
  {Park}}]{lee2022speed}%
  \BibitemOpen
  \bibfield  {author} {\bibinfo {author} {\bibfnamefont {J.~S.}\ \bibnamefont
  {Lee}}, \bibinfo {author} {\bibfnamefont {S.}~\bibnamefont {Lee}}, \bibinfo
  {author} {\bibfnamefont {H.}~\bibnamefont {Kwon}},\ and\ \bibinfo {author}
  {\bibfnamefont {H.}~\bibnamefont {Park}},\ }\bibfield  {title} {\bibinfo
  {title} {Speed limit for a highly irreversible process and tight finite-time
  landauer’s bound},\ }\href@noop {} {\bibfield  {journal} {\bibinfo
  {journal} {Physical review letters}\ }\textbf {\bibinfo {volume} {129}},\
  \bibinfo {pages} {120603} (\bibinfo {year} {2022})}\BibitemShut {NoStop}%
\bibitem [{\citenamefont {Falasco}\ and\ \citenamefont
  {Esposito}(2020)}]{falasco2020dissipation}%
  \BibitemOpen
  \bibfield  {author} {\bibinfo {author} {\bibfnamefont {G.}~\bibnamefont
  {Falasco}}\ and\ \bibinfo {author} {\bibfnamefont {M.}~\bibnamefont
  {Esposito}},\ }\bibfield  {title} {\bibinfo {title} {Dissipation-time
  uncertainty relation},\ }\href@noop {} {\bibfield  {journal} {\bibinfo
  {journal} {Physical Review Letters}\ }\textbf {\bibinfo {volume} {125}},\
  \bibinfo {pages} {120604} (\bibinfo {year} {2020})}\BibitemShut {NoStop}%
\bibitem [{\citenamefont {Van~Vu}\ and\ \citenamefont
  {Saito}(2023{\natexlab{b}})}]{van2023thermodynamic}%
  \BibitemOpen
  \bibfield  {author} {\bibinfo {author} {\bibfnamefont {T.}~\bibnamefont
  {Van~Vu}}\ and\ \bibinfo {author} {\bibfnamefont {K.}~\bibnamefont {Saito}},\
  }\bibfield  {title} {\bibinfo {title} {Thermodynamic unification of optimal
  transport: Thermodynamic uncertainty relation, minimum dissipation, and
  thermodynamic speed limits},\ }\href@noop {} {\bibfield  {journal} {\bibinfo
  {journal} {Physical Review X}\ }\textbf {\bibinfo {volume} {13}},\ \bibinfo
  {pages} {011013} (\bibinfo {year} {2023}{\natexlab{b}})}\BibitemShut
  {NoStop}%
\bibitem [{\citenamefont {Kuznets-Speck}\ and\ \citenamefont
  {Limmer}(2021)}]{kuznets2021dissipation}%
  \BibitemOpen
  \bibfield  {author} {\bibinfo {author} {\bibfnamefont {B.}~\bibnamefont
  {Kuznets-Speck}}\ and\ \bibinfo {author} {\bibfnamefont {D.~T.}\ \bibnamefont
  {Limmer}},\ }\bibfield  {title} {\bibinfo {title} {Dissipation bounds the
  amplification of transition rates far from equilibrium},\ }\href@noop {}
  {\bibfield  {journal} {\bibinfo  {journal} {Proceedings of the National
  Academy of Sciences}\ }\textbf {\bibinfo {volume} {118}},\ \bibinfo {pages}
  {e2020863118} (\bibinfo {year} {2021})}\BibitemShut {NoStop}%
\bibitem [{\citenamefont {Lynn}\ \emph
  {et~al.}(2022{\natexlab{a}})\citenamefont {Lynn}, \citenamefont {Holmes},
  \citenamefont {Bialek},\ and\ \citenamefont {Schwab}}]{arrowoftime}%
  \BibitemOpen
  \bibfield  {author} {\bibinfo {author} {\bibfnamefont {C.~W.}\ \bibnamefont
  {Lynn}}, \bibinfo {author} {\bibfnamefont {C.~M.}\ \bibnamefont {Holmes}},
  \bibinfo {author} {\bibfnamefont {W.}~\bibnamefont {Bialek}},\ and\ \bibinfo
  {author} {\bibfnamefont {D.~J.}\ \bibnamefont {Schwab}},\ }\bibfield  {title}
  {\bibinfo {title} {Decomposing the local arrow of time in interacting
  systems},\ }\href {https://doi.org/10.1103/PhysRevLett.129.118101} {\bibfield
   {journal} {\bibinfo  {journal} {Phys. Rev. Lett.}\ }\textbf {\bibinfo
  {volume} {129}},\ \bibinfo {pages} {118101} (\bibinfo {year}
  {2022}{\natexlab{a}})}\BibitemShut {NoStop}%
\bibitem [{\citenamefont {Lynn}\ \emph
  {et~al.}(2022{\natexlab{b}})\citenamefont {Lynn}, \citenamefont {Holmes},
  \citenamefont {Bialek},\ and\ \citenamefont {Schwab}}]{lynn2022emergence}%
  \BibitemOpen
  \bibfield  {author} {\bibinfo {author} {\bibfnamefont {C.~W.}\ \bibnamefont
  {Lynn}}, \bibinfo {author} {\bibfnamefont {C.~M.}\ \bibnamefont {Holmes}},
  \bibinfo {author} {\bibfnamefont {W.}~\bibnamefont {Bialek}},\ and\ \bibinfo
  {author} {\bibfnamefont {D.~J.}\ \bibnamefont {Schwab}},\ }\bibfield  {title}
  {\bibinfo {title} {Emergence of local irreversibility in complex interacting
  systems},\ }\href@noop {} {\bibfield  {journal} {\bibinfo  {journal}
  {Physical Review E}\ }\textbf {\bibinfo {volume} {106}},\ \bibinfo {pages}
  {034102} (\bibinfo {year} {2022}{\natexlab{b}})}\BibitemShut {NoStop}%
\bibitem [{\citenamefont {Ito}(2018)}]{ito2018stochastic}%
  \BibitemOpen
  \bibfield  {author} {\bibinfo {author} {\bibfnamefont {S.}~\bibnamefont
  {Ito}},\ }\bibfield  {title} {\bibinfo {title} {Stochastic thermodynamic
  interpretation of information geometry},\ }\href@noop {} {\bibfield
  {journal} {\bibinfo  {journal} {Physical review letters}\ }\textbf {\bibinfo
  {volume} {121}},\ \bibinfo {pages} {030605} (\bibinfo {year}
  {2018})}\BibitemShut {NoStop}%
\bibitem [{\citenamefont {Aurell}\ \emph {et~al.}(2011)\citenamefont {Aurell},
  \citenamefont {Mej{\'\i}a-Monasterio},\ and\ \citenamefont
  {Muratore-Ginanneschi}}]{aurell2011optimal}%
  \BibitemOpen
  \bibfield  {author} {\bibinfo {author} {\bibfnamefont {E.}~\bibnamefont
  {Aurell}}, \bibinfo {author} {\bibfnamefont {C.}~\bibnamefont
  {Mej{\'\i}a-Monasterio}},\ and\ \bibinfo {author} {\bibfnamefont
  {P.}~\bibnamefont {Muratore-Ginanneschi}},\ }\bibfield  {title} {\bibinfo
  {title} {Optimal protocols and optimal transport in stochastic
  thermodynamics},\ }\href@noop {} {\bibfield  {journal} {\bibinfo  {journal}
  {Physical review letters}\ }\textbf {\bibinfo {volume} {106}},\ \bibinfo
  {pages} {250601} (\bibinfo {year} {2011})}\BibitemShut {NoStop}%
\bibitem [{\citenamefont {Chennakesavalu}\ \emph {et~al.}(2023)\citenamefont
  {Chennakesavalu}, \citenamefont {Manikandan}, \citenamefont {Hu},\ and\
  \citenamefont {Rotskoff}}]{chennakesavalu2023adaptive}%
  \BibitemOpen
  \bibfield  {author} {\bibinfo {author} {\bibfnamefont {S.}~\bibnamefont
  {Chennakesavalu}}, \bibinfo {author} {\bibfnamefont {S.~K.}\ \bibnamefont
  {Manikandan}}, \bibinfo {author} {\bibfnamefont {F.}~\bibnamefont {Hu}},\
  and\ \bibinfo {author} {\bibfnamefont {G.~M.}\ \bibnamefont {Rotskoff}},\
  }\bibfield  {title} {\bibinfo {title} {Adaptive nonequilibrium design of
  actin-based metamaterials: fundamental and practical limits of control},\
  }\href@noop {} {\bibfield  {journal} {\bibinfo  {journal} {arXiv preprint
  arXiv:2306.10778}\ } (\bibinfo {year} {2023})}\BibitemShut {NoStop}%
\bibitem [{\citenamefont {Shiraishi}\ and\ \citenamefont
  {Saito}(2019)}]{shiraishi2019information}%
  \BibitemOpen
  \bibfield  {author} {\bibinfo {author} {\bibfnamefont {N.}~\bibnamefont
  {Shiraishi}}\ and\ \bibinfo {author} {\bibfnamefont {K.}~\bibnamefont
  {Saito}},\ }\bibfield  {title} {\bibinfo {title} {Information-theoretical
  bound of the irreversibility in thermal relaxation processes},\ }\href@noop
  {} {\bibfield  {journal} {\bibinfo  {journal} {Physical review letters}\
  }\textbf {\bibinfo {volume} {123}},\ \bibinfo {pages} {110603} (\bibinfo
  {year} {2019})}\BibitemShut {NoStop}%
\bibitem [{\citenamefont {Ch{\'e}trite}\ \emph {et~al.}(2021)\citenamefont
  {Ch{\'e}trite}, \citenamefont {Kumar},\ and\ \citenamefont
  {Bechhoefer}}]{chetrite2021metastable}%
  \BibitemOpen
  \bibfield  {author} {\bibinfo {author} {\bibfnamefont {R.}~\bibnamefont
  {Ch{\'e}trite}}, \bibinfo {author} {\bibfnamefont {A.}~\bibnamefont
  {Kumar}},\ and\ \bibinfo {author} {\bibfnamefont {J.}~\bibnamefont
  {Bechhoefer}},\ }\bibfield  {title} {\bibinfo {title} {The metastable mpemba
  effect corresponds to a non-monotonic temperature dependence of extractable
  work},\ }\href@noop {} {\bibfield  {journal} {\bibinfo  {journal} {arXiv
  preprint arXiv:2101.06394}\ } (\bibinfo {year} {2021})}\BibitemShut {NoStop}%
\bibitem [{\citenamefont {McClendon}\ \emph {et~al.}(2012)\citenamefont
  {McClendon}, \citenamefont {Hua}, \citenamefont {Barreiro},\ and\
  \citenamefont {Jacobson}}]{mcclendon2012comparing}%
  \BibitemOpen
  \bibfield  {author} {\bibinfo {author} {\bibfnamefont {C.~L.}\ \bibnamefont
  {McClendon}}, \bibinfo {author} {\bibfnamefont {L.}~\bibnamefont {Hua}},
  \bibinfo {author} {\bibfnamefont {G.}~\bibnamefont {Barreiro}},\ and\
  \bibinfo {author} {\bibfnamefont {M.~P.}\ \bibnamefont {Jacobson}},\
  }\bibfield  {title} {\bibinfo {title} {Comparing conformational ensembles
  using the kullback--leibler divergence expansion},\ }\href@noop {} {\bibfield
   {journal} {\bibinfo  {journal} {Journal of chemical theory and computation}\
  }\textbf {\bibinfo {volume} {8}},\ \bibinfo {pages} {2115} (\bibinfo {year}
  {2012})}\BibitemShut {NoStop}%
\bibitem [{\citenamefont {Galas}\ \emph {et~al.}(2017)\citenamefont {Galas},
  \citenamefont {Dewey}, \citenamefont {Kunert-Graf},\ and\ \citenamefont
  {Sakhanenko}}]{galas2017expansion}%
  \BibitemOpen
  \bibfield  {author} {\bibinfo {author} {\bibfnamefont {D.~J.}\ \bibnamefont
  {Galas}}, \bibinfo {author} {\bibfnamefont {G.}~\bibnamefont {Dewey}},
  \bibinfo {author} {\bibfnamefont {J.}~\bibnamefont {Kunert-Graf}},\ and\
  \bibinfo {author} {\bibfnamefont {N.~A.}\ \bibnamefont {Sakhanenko}},\
  }\bibfield  {title} {\bibinfo {title} {Expansion of the kullback-leibler
  divergence, and a new class of information metrics},\ }\href@noop {}
  {\bibfield  {journal} {\bibinfo  {journal} {Axioms}\ }\textbf {\bibinfo
  {volume} {6}},\ \bibinfo {pages} {8} (\bibinfo {year} {2017})}\BibitemShut
  {NoStop}%
\bibitem [{\citenamefont {Tritchler}\ \emph {et~al.}(2011)\citenamefont
  {Tritchler}, \citenamefont {Sucheston}, \citenamefont {Chanda},\ and\
  \citenamefont {Ramanathan}}]{tritchler2011information}%
  \BibitemOpen
  \bibfield  {author} {\bibinfo {author} {\bibfnamefont {D.~L.}\ \bibnamefont
  {Tritchler}}, \bibinfo {author} {\bibfnamefont {L.}~\bibnamefont
  {Sucheston}}, \bibinfo {author} {\bibfnamefont {P.}~\bibnamefont {Chanda}},\
  and\ \bibinfo {author} {\bibfnamefont {M.}~\bibnamefont {Ramanathan}},\
  }\bibfield  {title} {\bibinfo {title} {Information metrics in genetic
  epidemiology},\ }\href@noop {} {\bibfield  {journal} {\bibinfo  {journal}
  {Statistical applications in genetics and molecular biology}\ }\textbf
  {\bibinfo {volume} {10}} (\bibinfo {year} {2011})}\BibitemShut {NoStop}%
\bibitem [{\citenamefont {Lu}\ and\ \citenamefont
  {Raz}(2017)}]{lu2017nonequilibrium}%
  \BibitemOpen
  \bibfield  {author} {\bibinfo {author} {\bibfnamefont {Z.}~\bibnamefont
  {Lu}}\ and\ \bibinfo {author} {\bibfnamefont {O.}~\bibnamefont {Raz}},\
  }\bibfield  {title} {\bibinfo {title} {Nonequilibrium thermodynamics of the
  markovian mpemba effect and its inverse},\ }\href@noop {} {\bibfield
  {journal} {\bibinfo  {journal} {Proceedings of the National Academy of
  Sciences}\ }\textbf {\bibinfo {volume} {114}},\ \bibinfo {pages} {5083}
  (\bibinfo {year} {2017})}\BibitemShut {NoStop}%
\bibitem [{\citenamefont {Watanabe}(1960)}]{watanabe1960information}%
  \BibitemOpen
  \bibfield  {author} {\bibinfo {author} {\bibfnamefont {S.}~\bibnamefont
  {Watanabe}},\ }\bibfield  {title} {\bibinfo {title} {Information theoretical
  analysis of multivariate correlation},\ }\href@noop {} {\bibfield  {journal}
  {\bibinfo  {journal} {IBM Journal of research and development}\ }\textbf
  {\bibinfo {volume} {4}},\ \bibinfo {pages} {66} (\bibinfo {year}
  {1960})}\BibitemShut {NoStop}%
\bibitem [{\citenamefont {Somani}\ \emph {et~al.}(2009)\citenamefont {Somani},
  \citenamefont {Killian},\ and\ \citenamefont {Gilson}}]{somani2009sampling}%
  \BibitemOpen
  \bibfield  {author} {\bibinfo {author} {\bibfnamefont {S.}~\bibnamefont
  {Somani}}, \bibinfo {author} {\bibfnamefont {B.~J.}\ \bibnamefont
  {Killian}},\ and\ \bibinfo {author} {\bibfnamefont {M.~K.}\ \bibnamefont
  {Gilson}},\ }\bibfield  {title} {\bibinfo {title} {Sampling conformations in
  high dimensions using low-dimensional distribution functions},\ }\href@noop
  {} {\bibfield  {journal} {\bibinfo  {journal} {The Journal of chemical
  physics}\ }\textbf {\bibinfo {volume} {130}} (\bibinfo {year}
  {2009})}\BibitemShut {NoStop}%
\bibitem [{\citenamefont {Killian}\ \emph {et~al.}(2007)\citenamefont
  {Killian}, \citenamefont {Yundenfreund~Kravitz},\ and\ \citenamefont
  {Gilson}}]{killian2007extraction}%
  \BibitemOpen
  \bibfield  {author} {\bibinfo {author} {\bibfnamefont {B.~J.}\ \bibnamefont
  {Killian}}, \bibinfo {author} {\bibfnamefont {J.}~\bibnamefont
  {Yundenfreund~Kravitz}},\ and\ \bibinfo {author} {\bibfnamefont {M.~K.}\
  \bibnamefont {Gilson}},\ }\bibfield  {title} {\bibinfo {title} {Extraction of
  configurational entropy from molecular simulations via an expansion
  approximation},\ }\href@noop {} {\bibfield  {journal} {\bibinfo  {journal}
  {The Journal of chemical physics}\ }\textbf {\bibinfo {volume} {127}}
  (\bibinfo {year} {2007})}\BibitemShut {NoStop}%
\bibitem [{\citenamefont {Galas}\ and\ \citenamefont
  {Sakhanenko}(2016)}]{galas2016multivariate}%
  \BibitemOpen
  \bibfield  {author} {\bibinfo {author} {\bibfnamefont {D.~J.}\ \bibnamefont
  {Galas}}\ and\ \bibinfo {author} {\bibfnamefont {N.~A.}\ \bibnamefont
  {Sakhanenko}},\ }\bibfield  {title} {\bibinfo {title} {Multivariate
  information measures: a unification using m$\backslash$" obius operators on
  subset lattices},\ }\href@noop {} {\bibfield  {journal} {\bibinfo  {journal}
  {arXiv preprint arXiv:1601.06780}\ } (\bibinfo {year} {2016})}\BibitemShut
  {NoStop}%
\bibitem [{\citenamefont {Matsuda}(2000)}]{Matsuda}%
  \BibitemOpen
  \bibfield  {author} {\bibinfo {author} {\bibfnamefont {H.}~\bibnamefont
  {Matsuda}},\ }\bibfield  {title} {\bibinfo {title} {Physical nature of
  higher-order mutual information: Intrinsic correlations and frustration},\
  }\href {https://doi.org/10.1103/PhysRevE.62.3096} {\bibfield  {journal}
  {\bibinfo  {journal} {Phys. Rev. E}\ }\textbf {\bibinfo {volume} {62}},\
  \bibinfo {pages} {3096} (\bibinfo {year} {2000})}\BibitemShut {NoStop}%
\bibitem [{\citenamefont {Fenley}\ \emph {et~al.}(2014)\citenamefont {Fenley},
  \citenamefont {Killian}, \citenamefont {Hnizdo}, \citenamefont {Fedorowicz},
  \citenamefont {Sharp},\ and\ \citenamefont {Gilson}}]{fenley2014correlation}%
  \BibitemOpen
  \bibfield  {author} {\bibinfo {author} {\bibfnamefont {A.~T.}\ \bibnamefont
  {Fenley}}, \bibinfo {author} {\bibfnamefont {B.~J.}\ \bibnamefont {Killian}},
  \bibinfo {author} {\bibfnamefont {V.}~\bibnamefont {Hnizdo}}, \bibinfo
  {author} {\bibfnamefont {A.}~\bibnamefont {Fedorowicz}}, \bibinfo {author}
  {\bibfnamefont {D.~S.}\ \bibnamefont {Sharp}},\ and\ \bibinfo {author}
  {\bibfnamefont {M.~K.}\ \bibnamefont {Gilson}},\ }\bibfield  {title}
  {\bibinfo {title} {Correlation as a determinant of configurational entropy in
  supramolecular and protein systems},\ }\href@noop {} {\bibfield  {journal}
  {\bibinfo  {journal} {The Journal of Physical Chemistry B}\ }\textbf
  {\bibinfo {volume} {118}},\ \bibinfo {pages} {6447} (\bibinfo {year}
  {2014})}\BibitemShut {NoStop}%
\bibitem [{\citenamefont {Hough}\ and\ \citenamefont
  {Ou-Yang}(2002)}]{hough2002correlated}%
  \BibitemOpen
  \bibfield  {author} {\bibinfo {author} {\bibfnamefont {L.}~\bibnamefont
  {Hough}}\ and\ \bibinfo {author} {\bibfnamefont {H.}~\bibnamefont
  {Ou-Yang}},\ }\bibfield  {title} {\bibinfo {title} {Correlated motions of two
  hydrodynamically coupled particles confined in separate quadratic potential
  wells},\ }\href@noop {} {\bibfield  {journal} {\bibinfo  {journal} {Physical
  Review E}\ }\textbf {\bibinfo {volume} {65}},\ \bibinfo {pages} {021906}
  (\bibinfo {year} {2002})}\BibitemShut {NoStop}%
\bibitem [{\citenamefont {Kotar}\ \emph {et~al.}(2010)\citenamefont {Kotar},
  \citenamefont {Leoni}, \citenamefont {Bassetti}, \citenamefont
  {Lagomarsino},\ and\ \citenamefont {Cicuta}}]{kotar2010hydrodynamic}%
  \BibitemOpen
  \bibfield  {author} {\bibinfo {author} {\bibfnamefont {J.}~\bibnamefont
  {Kotar}}, \bibinfo {author} {\bibfnamefont {M.}~\bibnamefont {Leoni}},
  \bibinfo {author} {\bibfnamefont {B.}~\bibnamefont {Bassetti}}, \bibinfo
  {author} {\bibfnamefont {M.~C.}\ \bibnamefont {Lagomarsino}},\ and\ \bibinfo
  {author} {\bibfnamefont {P.}~\bibnamefont {Cicuta}},\ }\bibfield  {title}
  {\bibinfo {title} {Hydrodynamic synchronization of colloidal oscillators},\
  }\href@noop {} {\bibfield  {journal} {\bibinfo  {journal} {Proceedings of the
  National Academy of Sciences}\ }\textbf {\bibinfo {volume} {107}},\ \bibinfo
  {pages} {7669} (\bibinfo {year} {2010})}\BibitemShut {NoStop}%
\bibitem [{\citenamefont {Reichert}\ and\ \citenamefont
  {Stark}(2004)}]{reichert2004hydrodynamic}%
  \BibitemOpen
  \bibfield  {author} {\bibinfo {author} {\bibfnamefont {M.}~\bibnamefont
  {Reichert}}\ and\ \bibinfo {author} {\bibfnamefont {H.}~\bibnamefont
  {Stark}},\ }\bibfield  {title} {\bibinfo {title} {Hydrodynamic coupling of
  two rotating spheres trapped in harmonic potentials},\ }\href@noop {}
  {\bibfield  {journal} {\bibinfo  {journal} {Physical Review E}\ }\textbf
  {\bibinfo {volume} {69}},\ \bibinfo {pages} {031407} (\bibinfo {year}
  {2004})}\BibitemShut {NoStop}%
\bibitem [{\citenamefont {Paul}\ \emph {et~al.}(2018)\citenamefont {Paul},
  \citenamefont {Kumar},\ and\ \citenamefont {Banerjee}}]{paul2018two}%
  \BibitemOpen
  \bibfield  {author} {\bibinfo {author} {\bibfnamefont {S.}~\bibnamefont
  {Paul}}, \bibinfo {author} {\bibfnamefont {R.}~\bibnamefont {Kumar}},\ and\
  \bibinfo {author} {\bibfnamefont {A.}~\bibnamefont {Banerjee}},\ }\bibfield
  {title} {\bibinfo {title} {Two-point active microrheology in a viscous medium
  exploiting a motional resonance excited in dual-trap optical tweezers},\
  }\href@noop {} {\bibfield  {journal} {\bibinfo  {journal} {Physical Review
  E}\ }\textbf {\bibinfo {volume} {97}},\ \bibinfo {pages} {042606} (\bibinfo
  {year} {2018})}\BibitemShut {NoStop}%
\bibitem [{\citenamefont {Paul}\ \emph {et~al.}(2017)\citenamefont {Paul},
  \citenamefont {Laskar}, \citenamefont {Singh}, \citenamefont {Roy},
  \citenamefont {Adhikari},\ and\ \citenamefont {Banerjee}}]{paul2017direct}%
  \BibitemOpen
  \bibfield  {author} {\bibinfo {author} {\bibfnamefont {S.}~\bibnamefont
  {Paul}}, \bibinfo {author} {\bibfnamefont {A.}~\bibnamefont {Laskar}},
  \bibinfo {author} {\bibfnamefont {R.}~\bibnamefont {Singh}}, \bibinfo
  {author} {\bibfnamefont {B.}~\bibnamefont {Roy}}, \bibinfo {author}
  {\bibfnamefont {R.}~\bibnamefont {Adhikari}},\ and\ \bibinfo {author}
  {\bibfnamefont {A.}~\bibnamefont {Banerjee}},\ }\bibfield  {title} {\bibinfo
  {title} {Direct verification of the fluctuation-dissipation relation in
  viscously coupled oscillators},\ }\href@noop {} {\bibfield  {journal}
  {\bibinfo  {journal} {Physical Review E}\ }\textbf {\bibinfo {volume} {96}},\
  \bibinfo {pages} {050102} (\bibinfo {year} {2017})}\BibitemShut {NoStop}%
\bibitem [{\citenamefont {Doi}\ and\ \citenamefont
  {Edwards}(1988)}]{doi1988theory}%
  \BibitemOpen
  \bibfield  {author} {\bibinfo {author} {\bibfnamefont {M.}~\bibnamefont
  {Doi}}\ and\ \bibinfo {author} {\bibfnamefont {S.~F.}\ \bibnamefont
  {Edwards}},\ }\href@noop {} {\emph {\bibinfo {title} {The theory of polymer
  dynamics}}},\ Vol.~\bibinfo {volume} {73}\ (\bibinfo  {publisher} {oxford
  university press},\ \bibinfo {year} {1988})\BibitemShut {NoStop}%
\bibitem [{\citenamefont {Kumar}\ and\ \citenamefont
  {Bechhoefer}(2020)}]{kumar2020exponentially}%
  \BibitemOpen
  \bibfield  {author} {\bibinfo {author} {\bibfnamefont {A.}~\bibnamefont
  {Kumar}}\ and\ \bibinfo {author} {\bibfnamefont {J.}~\bibnamefont
  {Bechhoefer}},\ }\bibfield  {title} {\bibinfo {title} {Exponentially faster
  cooling in a colloidal system},\ }\href@noop {} {\bibfield  {journal}
  {\bibinfo  {journal} {Nature}\ }\textbf {\bibinfo {volume} {584}},\ \bibinfo
  {pages} {64} (\bibinfo {year} {2020})}\BibitemShut {NoStop}%
\bibitem [{\citenamefont {Bechhoefer}\ \emph {et~al.}(2021)\citenamefont
  {Bechhoefer}, \citenamefont {Kumar},\ and\ \citenamefont
  {Ch{\'e}trite}}]{bechhoefer2021fresh}%
  \BibitemOpen
  \bibfield  {author} {\bibinfo {author} {\bibfnamefont {J.}~\bibnamefont
  {Bechhoefer}}, \bibinfo {author} {\bibfnamefont {A.}~\bibnamefont {Kumar}},\
  and\ \bibinfo {author} {\bibfnamefont {R.}~\bibnamefont {Ch{\'e}trite}},\
  }\bibfield  {title} {\bibinfo {title} {A fresh understanding of the mpemba
  effect},\ }\href@noop {} {\bibfield  {journal} {\bibinfo  {journal} {Nature
  Reviews Physics}\ ,\ \bibinfo {pages} {1}} (\bibinfo {year}
  {2021})}\BibitemShut {NoStop}%
\bibitem [{\citenamefont {Biswas}\ and\ \citenamefont
  {Rajesh}(2023)}]{biswas2023mpemba}%
  \BibitemOpen
  \bibfield  {author} {\bibinfo {author} {\bibfnamefont {A.}~\bibnamefont
  {Biswas}}\ and\ \bibinfo {author} {\bibfnamefont {R.}~\bibnamefont
  {Rajesh}},\ }\bibfield  {title} {\bibinfo {title} {Mpemba effect for a
  brownian particle trapped in a single well potential},\ }\href@noop {}
  {\bibfield  {journal} {\bibinfo  {journal} {arXiv preprint arXiv:2305.06613}\
  } (\bibinfo {year} {2023})}\BibitemShut {NoStop}%
\bibitem [{\citenamefont {Deg{\"u}nther}\ and\ \citenamefont
  {Seifert}(2022)}]{degunther2022anomalous}%
  \BibitemOpen
  \bibfield  {author} {\bibinfo {author} {\bibfnamefont {J.}~\bibnamefont
  {Deg{\"u}nther}}\ and\ \bibinfo {author} {\bibfnamefont {U.}~\bibnamefont
  {Seifert}},\ }\bibfield  {title} {\bibinfo {title} {Anomalous relaxation from
  a non-equilibrium steady state: An isothermal analog of the mpemba effect},\
  }\href@noop {} {\bibfield  {journal} {\bibinfo  {journal} {Europhysics
  Letters}\ }\textbf {\bibinfo {volume} {139}},\ \bibinfo {pages} {41002}
  (\bibinfo {year} {2022})}\BibitemShut {NoStop}%
\bibitem [{\citenamefont {Ib{\'a}{\~n}ez}\ \emph {et~al.}(2024)\citenamefont
  {Ib{\'a}{\~n}ez}, \citenamefont {Dieball}, \citenamefont {Lasanta},
  \citenamefont {Godec},\ and\ \citenamefont {Rica}}]{ibanez2024heating}%
  \BibitemOpen
  \bibfield  {author} {\bibinfo {author} {\bibfnamefont {M.}~\bibnamefont
  {Ib{\'a}{\~n}ez}}, \bibinfo {author} {\bibfnamefont {C.}~\bibnamefont
  {Dieball}}, \bibinfo {author} {\bibfnamefont {A.}~\bibnamefont {Lasanta}},
  \bibinfo {author} {\bibfnamefont {A.}~\bibnamefont {Godec}},\ and\ \bibinfo
  {author} {\bibfnamefont {R.~A.}\ \bibnamefont {Rica}},\ }\bibfield  {title}
  {\bibinfo {title} {Heating and cooling are fundamentally asymmetric and
  evolve along distinct pathways},\ }\href@noop {} {\bibfield  {journal}
  {\bibinfo  {journal} {Nature Physics}\ ,\ \bibinfo {pages} {1}} (\bibinfo
  {year} {2024})}\BibitemShut {NoStop}%
\bibitem [{\citenamefont {Lapolla}\ and\ \citenamefont {Godec}(2020)}]{Uphill}%
  \BibitemOpen
  \bibfield  {author} {\bibinfo {author} {\bibfnamefont {A.}~\bibnamefont
  {Lapolla}}\ and\ \bibinfo {author} {\bibfnamefont {A.~c.~v.}\ \bibnamefont
  {Godec}},\ }\bibfield  {title} {\bibinfo {title} {Faster uphill relaxation in
  thermodynamically equidistant temperature quenches},\ }\href
  {https://doi.org/10.1103/PhysRevLett.125.110602} {\bibfield  {journal}
  {\bibinfo  {journal} {Phys. Rev. Lett.}\ }\textbf {\bibinfo {volume} {125}},\
  \bibinfo {pages} {110602} (\bibinfo {year} {2020})}\BibitemShut {NoStop}%
\bibitem [{\citenamefont {Manikandan}(2021)}]{fewlevelQ}%
  \BibitemOpen
  \bibfield  {author} {\bibinfo {author} {\bibfnamefont {S.~K.}\ \bibnamefont
  {Manikandan}},\ }\bibfield  {title} {\bibinfo {title} {Equidistant quenches
  in few-level quantum systems},\ }\href
  {https://doi.org/10.1103/PhysRevResearch.3.043108} {\bibfield  {journal}
  {\bibinfo  {journal} {Phys. Rev. Res.}\ }\textbf {\bibinfo {volume} {3}},\
  \bibinfo {pages} {043108} (\bibinfo {year} {2021})}\BibitemShut {NoStop}%
\bibitem [{\citenamefont {Van~Vu}\ and\ \citenamefont
  {Hasegawa}(2021)}]{van2021toward}%
  \BibitemOpen
  \bibfield  {author} {\bibinfo {author} {\bibfnamefont {T.}~\bibnamefont
  {Van~Vu}}\ and\ \bibinfo {author} {\bibfnamefont {Y.}~\bibnamefont
  {Hasegawa}},\ }\bibfield  {title} {\bibinfo {title} {Toward relaxation
  asymmetry: Heating is faster than cooling},\ }\href@noop {} {\bibfield
  {journal} {\bibinfo  {journal} {Physical Review Research}\ }\textbf {\bibinfo
  {volume} {3}},\ \bibinfo {pages} {043160} (\bibinfo {year}
  {2021})}\BibitemShut {NoStop}%
\bibitem [{\citenamefont {Dieball}\ \emph {et~al.}(2023)\citenamefont
  {Dieball}, \citenamefont {Wellecke},\ and\ \citenamefont
  {Godec}}]{dieball2023asymmetric}%
  \BibitemOpen
  \bibfield  {author} {\bibinfo {author} {\bibfnamefont {C.}~\bibnamefont
  {Dieball}}, \bibinfo {author} {\bibfnamefont {G.}~\bibnamefont {Wellecke}},\
  and\ \bibinfo {author} {\bibfnamefont {A.}~\bibnamefont {Godec}},\ }\bibfield
   {title} {\bibinfo {title} {Asymmetric thermal relaxation in driven systems:
  Rotations go opposite ways},\ }\href@noop {} {\bibfield  {journal} {\bibinfo
  {journal} {arXiv preprint arXiv:2304.06702}\ } (\bibinfo {year}
  {2023})}\BibitemShut {NoStop}%
\bibitem [{\citenamefont {Meibohm}\ \emph {et~al.}(2021)\citenamefont
  {Meibohm}, \citenamefont {Forastiere}, \citenamefont {Adeleke-Larodo},\ and\
  \citenamefont {Proesmans}}]{meibohm2021relaxation}%
  \BibitemOpen
  \bibfield  {author} {\bibinfo {author} {\bibfnamefont {J.}~\bibnamefont
  {Meibohm}}, \bibinfo {author} {\bibfnamefont {D.}~\bibnamefont {Forastiere}},
  \bibinfo {author} {\bibfnamefont {T.}~\bibnamefont {Adeleke-Larodo}},\ and\
  \bibinfo {author} {\bibfnamefont {K.}~\bibnamefont {Proesmans}},\ }\bibfield
  {title} {\bibinfo {title} {Relaxation-speed crossover in anharmonic
  potentials},\ }\href@noop {} {\bibfield  {journal} {\bibinfo  {journal}
  {Physical Review E}\ }\textbf {\bibinfo {volume} {104}},\ \bibinfo {pages}
  {L032105} (\bibinfo {year} {2021})}\BibitemShut {NoStop}%
\bibitem [{\citenamefont {Ghimenti}\ \emph {et~al.}(2023)\citenamefont
  {Ghimenti}, \citenamefont {Berthier}, \citenamefont {Szamel},\ and\
  \citenamefont {van Wijland}}]{ghimenti2023sampling}%
  \BibitemOpen
  \bibfield  {author} {\bibinfo {author} {\bibfnamefont {F.}~\bibnamefont
  {Ghimenti}}, \bibinfo {author} {\bibfnamefont {L.}~\bibnamefont {Berthier}},
  \bibinfo {author} {\bibfnamefont {G.}~\bibnamefont {Szamel}},\ and\ \bibinfo
  {author} {\bibfnamefont {F.}~\bibnamefont {van Wijland}},\ }\bibfield
  {title} {\bibinfo {title} {Sampling efficiency of transverse forces in dense
  liquids},\ }\href@noop {} {\bibfield  {journal} {\bibinfo  {journal}
  {Physical Review Letters}\ }\textbf {\bibinfo {volume} {131}},\ \bibinfo
  {pages} {257101} (\bibinfo {year} {2023})}\BibitemShut {NoStop}%
\bibitem [{\citenamefont {Tang}\ \emph {et~al.}(2015)\citenamefont {Tang},
  \citenamefont {Yuan}, \citenamefont {Chen},\ and\ \citenamefont {Ao}}]{Tang}%
  \BibitemOpen
  \bibfield  {author} {\bibinfo {author} {\bibfnamefont {Y.}~\bibnamefont
  {Tang}}, \bibinfo {author} {\bibfnamefont {R.}~\bibnamefont {Yuan}}, \bibinfo
  {author} {\bibfnamefont {J.}~\bibnamefont {Chen}},\ and\ \bibinfo {author}
  {\bibfnamefont {P.}~\bibnamefont {Ao}},\ }\bibfield  {title} {\bibinfo
  {title} {Work relations connecting nonequilibrium steady states without
  detailed balance},\ }\href {https://doi.org/10.1103/PhysRevE.91.042108}
  {\bibfield  {journal} {\bibinfo  {journal} {Phys. Rev. E}\ }\textbf {\bibinfo
  {volume} {91}},\ \bibinfo {pages} {042108} (\bibinfo {year}
  {2015})}\BibitemShut {NoStop}%
\bibitem [{Note1()}]{Note1}%
  \BibitemOpen
  \bibinfo {note} {See Supplemental Material at [URL will be inserted by
  publisher] for a basic implementation of the decomposition of the distance
  function in python.}\BibitemShut {Stop}%
\bibitem [{\citenamefont {Yan}\ \emph {et~al.}(2022)\citenamefont {Yan},
  \citenamefont {Touchette}, \citenamefont {Rotskoff} \emph
  {et~al.}}]{yan2022learning}%
  \BibitemOpen
  \bibfield  {author} {\bibinfo {author} {\bibfnamefont {J.}~\bibnamefont
  {Yan}}, \bibinfo {author} {\bibfnamefont {H.}~\bibnamefont {Touchette}},
  \bibinfo {author} {\bibfnamefont {G.~M.}\ \bibnamefont {Rotskoff}}, \emph
  {et~al.},\ }\bibfield  {title} {\bibinfo {title} {Learning nonequilibrium
  control forces to characterize dynamical phase transitions},\ }\href@noop {}
  {\bibfield  {journal} {\bibinfo  {journal} {Physical Review E}\ }\textbf
  {\bibinfo {volume} {105}},\ \bibinfo {pages} {024115} (\bibinfo {year}
  {2022})}\BibitemShut {NoStop}%
\bibitem [{\citenamefont {Rotskoff}\ \emph {et~al.}(2017)\citenamefont
  {Rotskoff}, \citenamefont {Crooks},\ and\ \citenamefont
  {Vanden-Eijnden}}]{rotskoff2017geometric}%
  \BibitemOpen
  \bibfield  {author} {\bibinfo {author} {\bibfnamefont {G.~M.}\ \bibnamefont
  {Rotskoff}}, \bibinfo {author} {\bibfnamefont {G.~E.}\ \bibnamefont
  {Crooks}},\ and\ \bibinfo {author} {\bibfnamefont {E.}~\bibnamefont
  {Vanden-Eijnden}},\ }\bibfield  {title} {\bibinfo {title} {Geometric approach
  to optimal nonequilibrium control: Minimizing dissipation in nanomagnetic
  spin systems},\ }\href@noop {} {\bibfield  {journal} {\bibinfo  {journal}
  {Physical Review E}\ }\textbf {\bibinfo {volume} {95}},\ \bibinfo {pages}
  {012148} (\bibinfo {year} {2017})}\BibitemShut {NoStop}%
\bibitem [{\citenamefont {Abreu}\ and\ \citenamefont
  {Seifert}(2012)}]{abreu2012thermodynamics}%
  \BibitemOpen
  \bibfield  {author} {\bibinfo {author} {\bibfnamefont {D.}~\bibnamefont
  {Abreu}}\ and\ \bibinfo {author} {\bibfnamefont {U.}~\bibnamefont
  {Seifert}},\ }\bibfield  {title} {\bibinfo {title} {Thermodynamics of genuine
  nonequilibrium states under feedback control},\ }\href@noop {} {\bibfield
  {journal} {\bibinfo  {journal} {Physical review letters}\ }\textbf {\bibinfo
  {volume} {108}},\ \bibinfo {pages} {030601} (\bibinfo {year}
  {2012})}\BibitemShut {NoStop}%
\bibitem [{\citenamefont {Gour}\ \emph {et~al.}(2015)\citenamefont {Gour},
  \citenamefont {M{\"u}ller}, \citenamefont {Narasimhachar}, \citenamefont
  {Spekkens},\ and\ \citenamefont {Halpern}}]{gour2015resource}%
  \BibitemOpen
  \bibfield  {author} {\bibinfo {author} {\bibfnamefont {G.}~\bibnamefont
  {Gour}}, \bibinfo {author} {\bibfnamefont {M.~P.}\ \bibnamefont
  {M{\"u}ller}}, \bibinfo {author} {\bibfnamefont {V.}~\bibnamefont
  {Narasimhachar}}, \bibinfo {author} {\bibfnamefont {R.~W.}\ \bibnamefont
  {Spekkens}},\ and\ \bibinfo {author} {\bibfnamefont {N.~Y.}\ \bibnamefont
  {Halpern}},\ }\bibfield  {title} {\bibinfo {title} {The resource theory of
  informational nonequilibrium in thermodynamics},\ }\href@noop {} {\bibfield
  {journal} {\bibinfo  {journal} {Physics Reports}\ }\textbf {\bibinfo {volume}
  {583}},\ \bibinfo {pages} {1} (\bibinfo {year} {2015})}\BibitemShut {NoStop}%
\bibitem [{\citenamefont {Argun}\ \emph {et~al.}(2017)\citenamefont {Argun},
  \citenamefont {Soni}, \citenamefont {Dabelow}, \citenamefont {Bo},
  \citenamefont {Pesce}, \citenamefont {Eichhorn},\ and\ \citenamefont
  {Volpe}}]{aykut}%
  \BibitemOpen
  \bibfield  {author} {\bibinfo {author} {\bibfnamefont {A.}~\bibnamefont
  {Argun}}, \bibinfo {author} {\bibfnamefont {J.}~\bibnamefont {Soni}},
  \bibinfo {author} {\bibfnamefont {L.}~\bibnamefont {Dabelow}}, \bibinfo
  {author} {\bibfnamefont {S.}~\bibnamefont {Bo}}, \bibinfo {author}
  {\bibfnamefont {G.}~\bibnamefont {Pesce}}, \bibinfo {author} {\bibfnamefont
  {R.}~\bibnamefont {Eichhorn}},\ and\ \bibinfo {author} {\bibfnamefont
  {G.}~\bibnamefont {Volpe}},\ }\bibfield  {title} {\bibinfo {title}
  {Experimental realization of a minimal microscopic heat engine},\ }\href
  {https://doi.org/10.1103/PhysRevE.96.052106} {\bibfield  {journal} {\bibinfo
  {journal} {Phys. Rev. E}\ }\textbf {\bibinfo {volume} {96}},\ \bibinfo
  {pages} {052106} (\bibinfo {year} {2017})}\BibitemShut {NoStop}%
\end{thebibliography}

%

\end{document}